\documentclass[12pt]{article}  
 
\usepackage{epsfig} 
\usepackage{amsmath,amssymb,amsfonts} 
\usepackage{float,a4wide} 
\usepackage{cite,slashed}
\usepackage{bbm,color}
\newcommand{\be}{\begin{equation}} 
\newcommand{\en}{\end{equation}}
\newcommand{\bea}{\begin{eqnarray}}
\newcommand{\ena}{\end{eqnarray}}

\newcommand{\hbo}{\hbox to 1 true cm {\hfill } } 
\newcommand{\tr}{\hbox{tr}}

\newcommand{\ft}[2]{{\textstyle\frac{#1}{#2}}}
\newcommand{\mbf}[1]{\boldmathe{#1}}
\DeclareMathAlphabet{\boldmathe}{T1}{cmr}{bx}{it}
\def\one{\mathbbm{1}}
\def\Z{\mathbbm{ Z}}
\def\dslash{\partial\kern-.5em\slash}
\def\kslash{k\kern-.5em\slash}
\def\pslash{p\kern-.5em\slash}
\def\Dslash{D\kern-.5em\slash}
\def\fdirac{\slashed{\partial}}

\def\ha{\frac{1}{2}}
\def\R{{\mathbb{R}}}
\def\E{\mathrm{e}}
\def\I{{\rm i}}
\def\cP{\mathcal{P}}
\def\mbx{\mbf{x}}
\def\mby{\mbf{y}}

\begin{document} 
\vglue 1truecm
  
\vbox{ 
\hfill 2 September 2011 
}
  
\vfil
\centerline{\large\bf Fermi-Einstein condensation in dense QCD-like theories }

\bigskip\bigskip
\bigskip\bigskip
\centerline{ Kurt Langfeld$^a$, Andreas Wipf$^b$ }
\vspace{.5 true cm} 
\centerline{$^a$ School of Computing \& Mathematics, University of Plymouth }
\centerline{Plymouth, PL4 8AA, UK }
  
\vspace{.5 true cm} 
\centerline{$^b$Theoretisch-Physikalisches Institut, Friedrich-Schiller-Universit\"at Jena  }
\centerline{07743 Jena, Germany}

\vskip 1.5cm

\begin{abstract}
While pure Yang-Mills theory feature the centre symmetry, this symmetry 
is explicitly broken by the presence of dynamical matter. We study the impact
of  the centre symmetry in such QCD-like theories. 
In the analytically solvable Schwinger model, centre transitions take place 
even under extreme conditions, temperature and/or density, and we show 
that they are key to the solution of the Silver-Blaze problem. 
We then develop an effective SU(3) quark model which confines 
quarks by virtue of centre sector transitions. The phase diagram by 
confinement is obtained as a function of the temperature and the chemical
potential. We show that at low temperatures and intermediate
values for the chemical potential the centre dressed quarks undergo
condensation due to Bose like statistics. 
This is the \emph{Fermi Einstein condensation}. 
To corroborate the existence of centre sector transitions 
in gauge theories with matter, we study (at vanishing chemical potential)  the
interface 
tension in the three-dimensional $\Z_2$ gauge theory with Ising matter,
the distribution of the Polyakov line in the four-dimensional SU(2)-Higgs
model and devise a new type of order parameter which is designed to detect
centre sector transitions. Our analytical and numerical findings lead us to
conjecture a new state of cold, but dense matter in the hadronic
phase for which Fermi Einstein condensation is realised.
%
%
\end{abstract}

\vfil
\hrule width 5truecm
\vskip .2truecm
\begin{quote} 
\end{quote}
\eject

\tableofcontents

\section{Introduction:}

A great deal of efforts, theoretical and experimental, are devoted 
to explore the properties of hadronic matter under extreme conditions, 
temperature and/or baryonic densities. The general belief is that 
matter is organised in a quark-gluon plasma phase in this regime. 
At least for small densities, this belief is corroborated by 
lattice gauge calculations~\cite{Karsch:2001cy,Boyd:1996bx,Cheng:2007jq}
and collision experiments such as undertaken at
RHIC~\cite{Shuryak:2008eq,Frawley:2008kk}.  

\vskip 0.3cm 
Central to understand matter under extreme conditions is the understanding 
of colour confinement since its realisation is, almost by definition, the most 
prominent difference between the so-called hadronic and the quark-gluon 
plasma phase. Early on in the eighties, it was pointed out that the 
centre of the gauge group plays a major role for confinement: 
Realisation of the centre symmetry of the gauge sector goes in line 
with confinement while its spontaneous breakdown (e.g.~at high temperatures) 
signals colour liberation~\cite{Svetitsky:1982gs,Svetitsky:1985ye}. 
Roughly at the same time, it was proposed that certain degrees of freedom 
of Yang-Mills theories, such as monopoles or vortices, are responsible 
for the (dis-)order of the centre symmetry (see~\cite{Greensite:2003bk}
for a review). It took until the late nineties to isolate 
these degrees of freedom in lattice gauge theories in a physical, i.e., 
regulator independent way~\cite{DelDebbio:1998uu,Langfeld:1997jx}. 
It was pointed out in~\cite{Zakharov:2006vt} that the mere existence 
of centre vortices as physical degrees of freedom demand a fine tuning 
between vortex entropy and energy. Quite recently~\cite{Langfeld:2010nm}, 
smooth stream-line configurations which bear confinement have been found using
lattice gauge simulations: it was pointed out that if centre vortices are
images of these 
stream-line configurations in a certain gauge, it would naturally 
account for the intrinsic fine-tuning. Most importantly, 
the centre vortex picture offers an understanding 
of confinement on the basis of weakly interacting degrees of 
freedom, which explain high temperature deconfinement and the 
centre disorder of spatial correlations at the same 
time~\cite{Langfeld:1998cz,Engelhardt:1999fd,Langfeld:2003zi}. 

\vskip 0.3cm 
Although a quite detailed picture of confinement and deconfinement 
at finite temperatures has emerged over the last decade, very little is 
known about cold but dense matter from first principles. 
Because of the notorious ``sign''-problem, Monte-Carlo simulations 
of Yang-Mills theories with dense quark matter are only feasible 
for SU(2) colour~\cite{Kogut:2001na} for which the sign problem 
is absent\footnote{$G_2$ Yang-Mills theory with dynamical
fermions has no sign-problem as well and simulations at finite
temperature and density are under way~\cite{Wellegehausen:2011}.}
or for rather small densities (see~\cite{deForcrand:2010ys} for 
a recent review). First insights into the properties of matter which 
feature in cold but dense QCD might be gained from exact solutions 
of models which mimic certain aspects of QCD. As an example, we mention 
the Gross-Neveu model which features the baryon crystal as an hitherto 
unknown state of matter~\cite{Schnetz:2004vr,Schnetz:2005ih,Boehmer:2007ea}. 

\vskip 0.3cm 
Among the very few gauge theories which admit an exact solution even 
for the case of dense quark matter is 2d Quantumelectrodynamics, the so-called
Schwinger model~\cite{Schwinger:1962tp}. The model with massless fermions was
exactly solved in Hamiltonian formalism on the line 
in~\cite{Brown:1963,Lowenstein:1971fc,Casher:1974vf}
and on $S^1$ in \cite{Manton:1985jm,Iso:1988zi}: chiral symmetry is 
spontaneously broken and only states with a vanishing net baryon number 
appear in the spectrum. 
The model on the torus has been studied in \cite{Joos:1990km} and in 
particular the temperature dependence  of the chiral condensate, 
Wilson loop correlators and Polyakov line correlators have been 
determined~\cite{Sachs:1992pa,Smilga:1992hx,Azakov:1996xk}. 
In~\cite{Sachs:1995dm,AlvarezEstrada:1997ja}, non-vanishing values of the 
fermion chemical potential have been firstly considered. It was found, 
most importantly, that the full non-perturbative partition function is independent 
of the chemical potential. Since then, many more quantities have been 
obtained in the massless Schwinger model ranging from the 
temperature dependence of the correlators of hadronic currents, 
spectral functions to the screening mass \cite{Fayyazuddin:1993ua}.
This makes the Schwinger mode the ideal testbed to test new ideas, and 
we will make extensive use of this below. 

\vskip 0.3cm 
Besides of exactly solvable models, understanding {\it mechanisms}, which have 
been revealed in model studies or by non-perturbative approximations, 
is an invaluable tool since they might extend their applicability 
to the theory of interest, in particular, cold and dense QCD matter. 
Recent examples are the proposal of the {\it quarkyonic} phase the existence 
of which has been motivated by the large $N_c$ considerations ($N_c$ 
being the number of colours)~\cite{McLerran:2007qj,McLerran:2008ua}. 
Another recent example is the chiral magnetic effect
which describes an induced electromagnetic current alongside an 
external magnetic field made possible by topological charge transitions 
in the quark gluon plasma phase~\cite{Fukushima:2008xe,Buividovich:2009wi}. 

\vskip 0.3cm 
Of particular importance for this publication is the 
recent observation that quarks effectively comply to {\it periodic 
boundaries conditions} if exposed to a particular non-trivial 
centre sector in $SU(N_c)$ Yang-Mills theory for $N_c$ even. 
In the dense hadronic (confining) phase, this opens the possibility that 
centre dressed quarks undergo condensation reminiscent of 
Bose-Einstein condensation~\cite{Langfeld:2009cd}. 
In analogy, this has been called {\it Fermi Einstein 
condensation (FEC)}~\cite{Langfeldect2010}. We also point out that the
sensitivity of  the quark spectrum to the boundary conditions of the quarks 
has been studied in~\cite{Bilgici:2009tx}.

\vskip 0.3cm 
In this paper, we will further study the phenomenological 
impact of the FEC effect. We will use the Schwinger model 
which will allow to study FEC on almost purely analytical grounds. 
Since in previous studies we considered an even number of colours, 
we here extend the considerations to the more relevant case 
of an $SU(N_c=3)$ gauge group using a quark model. We are going to show 
that transitions between the centre sectors are sufficient to 
confine quarks in this model. The quark model itself determines 
the so-called centre sector weights which serve as an order 
parameter for confinement. Using these weights, we will be able 
to calculate the phase diagram of the model as a function of the 
chemical potential and the temperature. For $N_c=3$, we find 
that FEC does occur under cold and dense conditions at the presence of
pressure. Essential for FEC is that the gluonic states manoeuvre 
through the centre sectors. Since non-trivial centre-sectors are suppressed 
by the presence of dynamical matter, we accumulate evidence in the 
remainder of the paper that centre sector transitions do occur 
in the so-called hadronic phase. To this aim, we will study 
the $\Z_2$ gauge theory with Ising matter and the SU(2) Higgs theory. 
We develop a novel order parameter which is sensitive to centre sector 
transitions and provide numerical evidence that these transitions 
take place until centre symmetry is spontaneously broken at 
high temperatures. 

\vskip 0.3cm 
The paper is organised as follows: In section~\ref{sec:YMmod}, we 
extend the considerations
of~\cite{Keurentjes:1998uu,Selivanov:1999cr,Schaden:2004ah} and show 
that the Yang-Mills ``empty'' vacuum possess gauge in-equivalent flat
directions  which collapse to the so-called centre-sectors once quantum 
fluctuations are included. Section~\ref{sec:Schw} addresses FEC in 
the context of the Schwinger model. In section~\ref{sec:Qconf}, we 
develop the $SU(3)$ quark model with confinement by virtue 
of the interaction of the quarks with the centre sector background 
fields. Confinement is established by studying the model's thermal 
excitations, and the phase diagram from confinement is obtained. 
In section~\ref{sec:gi}, we start to investigate centre-sector transition 
by means of numerical simulations. For this purpose, we 
study the $\Z_2$ gauge theory with Ising matter, which has not yet 
been studied before to our knowledge. In section~\ref{sec:higgs}, 
we extend the simulations and consider the SU(2) Higgs theory. 
The order parameter for centre sector transitions is developed and 
our numerical findings for this order parameter are presented. 
Conclusions are left to the final section.

\section{Yang-Mills moduli space and centre sectors \label{sec:YMmod} }

\subsection{The empty vacuum on a torus \label{sec:empty} } 

A configuration with minimal Euclidean action, often called {\it empty} vacuum, 
is often the starting point of perturbation theory. In Abelian theories, 
it is naturally defined as a state for which the field strength at any 
point of space-time vanishes. In non-Abelian theories, such as $SU(N_c)$
Yang-Mills theories, a more stringent definition is in order: there, the 
empty  vacuum is a state for which the holonomy calculated along 
any contractible loop ${\cal C}$ yields the unit element of the 
gauge group: 
\be 
\mathrm{P} \; \exp \left\{ \I \int _{\cal C} A_\mu(x) \, dx^\mu \right\} 
\; = \; \one  \hbo \hbox{(empty vacuum condition)} , 
\label{eq:pv1}
\en 
where $A_\mu(x)$ is the gauge potential and $P$ denotes path-ordering. 
Throughout this paper, we will consider a 4-torus as space-time 
manifold. The gauge potentials $A_\mu (x)$  and gauge transformations 
$\Omega (x)$ satisfy periodic boundary conditions\footnote{In the
continuum theory this implies a vanishing instanton number.},
matter fields such as a scalar Higgs field $\phi (x)$ or quarks $q(x)$ are subjected 
to periodic and anti-periodic boundary conditions, respectively. 
We will adopt a lattice regularisation with the lattice spacing $a$ 
acting as an UV regulator. Thereby, the gauge degrees of freedom 
are represented by the links $U_\mu (x) \in SU(N_c)$. Gauge transformations 
act as usual: 
\be 
U^\Omega _\mu (x) \; = \; \Omega (x) \, U_\mu (x) \, \Omega ^\dagger 
(x+\mu) \; . 
\label{eq:pv2}
\en
In lattice discretisation, the vacuum condition (\ref{eq:pv1}) 
becomes 
\be 
\prod _{\ell\in {\cal C}} U_\ell \; = \; \one \; , \hbo 
\ell = (x,\mu) \; , 
\label{eq:pv3}
\en 
where ${\cal C}$ is a closed and {\it contractible} path on the 
lattice. 
The smallest contractible loops on the lattice are the loops 
surrounding the elementary plaquettes of the lattice. 
For the later choice, the vacuum condition (\ref{eq:pv3}) implies 
that the non-Abelian field strength vanishes for the empty vacuum. 

\vskip 0.3cm 
Already in the late nineties, it has been discovered that 
there are smoothly connected configurations which are gauge {\it in-equivalent} 
and which all satisfy the vacuum condition (\ref{eq:pv1}) (or 
(\ref{eq:pv3}))~\cite{Keurentjes:1998uu,Selivanov:1999cr,Schaden:2004ah}. 
The set of these configurations defines the Yang-Mills moduli space. 
While Keurentjes et~al.~discuss 3-dimensional Yang-Mills theory, 
the arguments were extended to 4-dimension in~\cite{Schaden:2004ah}. 
To make the paper self-contained, we will give a full account of 
the Yang-Mills moduli space using lattice regularisation. 

\vskip 0.3cm 
\begin{figure}
  \includegraphics[width=6.5cm]{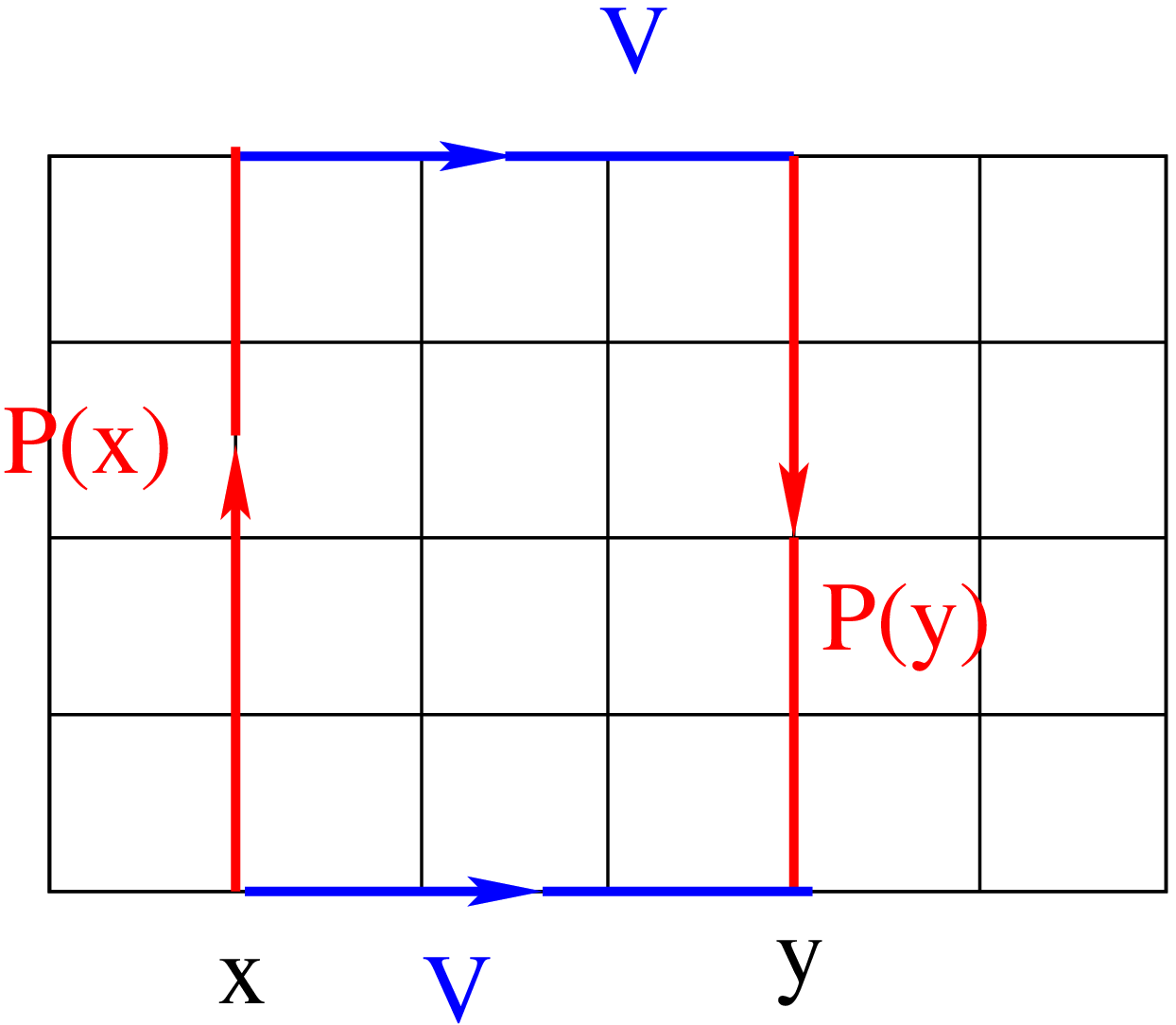} \hspace{0.5cm}
  \includegraphics[height=5cm]{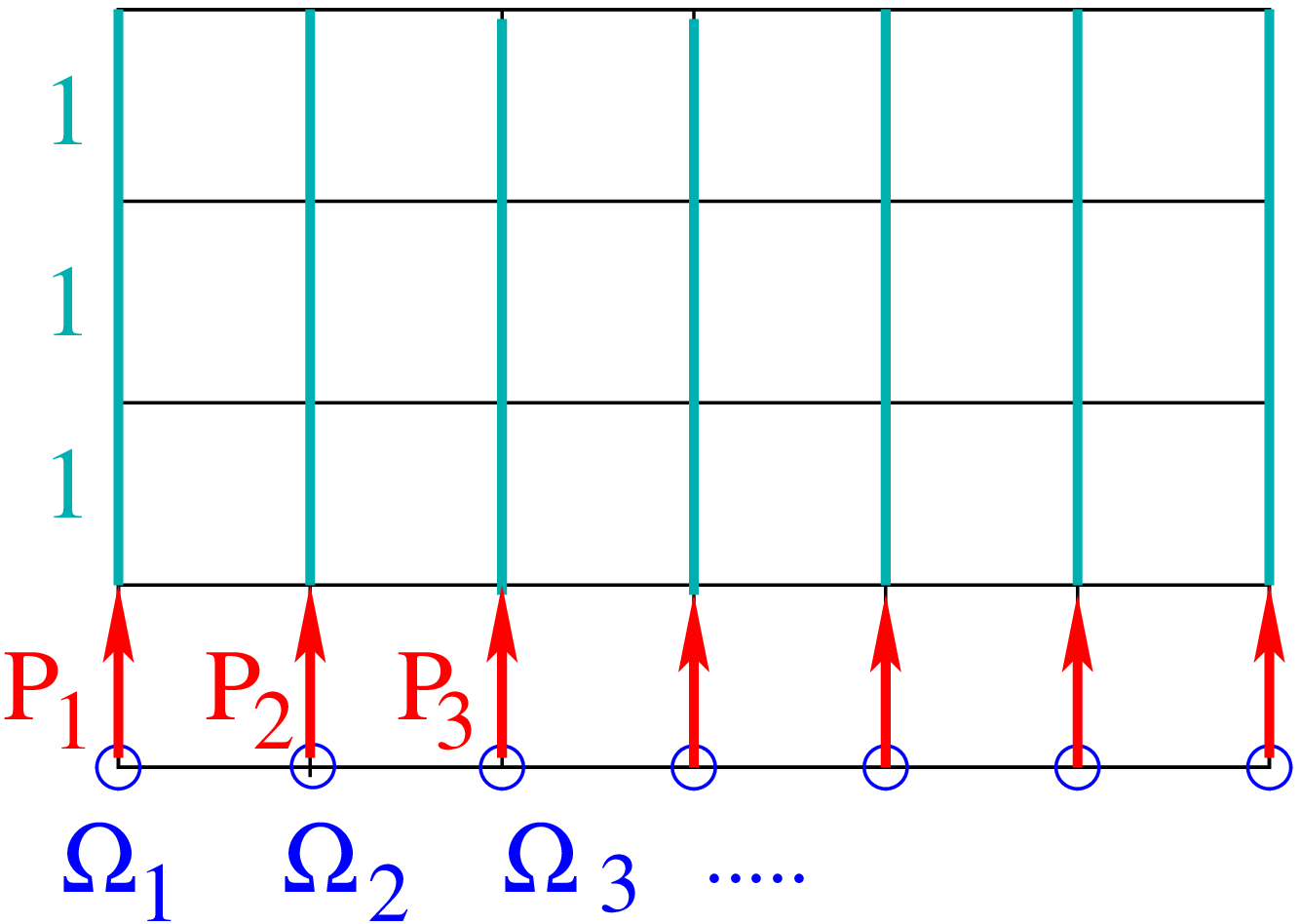}
\caption{\label{fig:1} Absence of Polyakov line correlations in 
the ``empty'' vacuum (left). Step 1 of the complete gauge fixing (right).
}
\end{figure}
Of particular importance for the study of confinement in Yang-Mills 
theories on the torus is the Polyakov line: 
\be 
\cP(\mbx) \; = \; \prod _{t} U_0(t,\mbx) \; . 
\label{eq:pv4}
\en 
Let us firstly study their correlations in the empty vacuum. For 
this purpose, we consider the maximal (contractible) loop in 
figure~\ref{fig:1}, left panel (note that the product $V$ of spatial links 
is the same at the lower and upper time-slices due 
to periodic boundary conditions). The vacuum condition implies 
\be 
\cP(\mbx)  V  \cP^\dagger(\mby)  V^\dagger \; = \; \one \hbo \Rightarrow \hbo 
\tr \, \cP(\mbx) \; = \; \tr \, \cP(\mby) . 
\label{eq:pv5}
\en 
Hence, the trace of the Polyakov line is necessarily constant for 
any choice of an ``empty'' vacuum. Given that the static quark 
anti-quark potential $V(r)$ can be extracted from the Polyakov line 
correlator: 
\be 
\bigl\langle P(\mbx) \, P(\mby) \bigr\rangle 
\propto \exp \left\{ - V(r) /T \right\}, \hbo 
P(\mbx) = \frac{1}{N_c} \, \tr \, \cP(\mbx), 
\label{eq:pv6}
\en 
where $r=\vert \mbx-\mby\vert$ and $T$ is the temperature, 
the finding~(\ref{eq:pv6}) implies that a single ``empty'' state 
cannot sustain quark confinement. 

\vskip 0.3cm 
\begin{figure}
  \includegraphics[width=7cm]{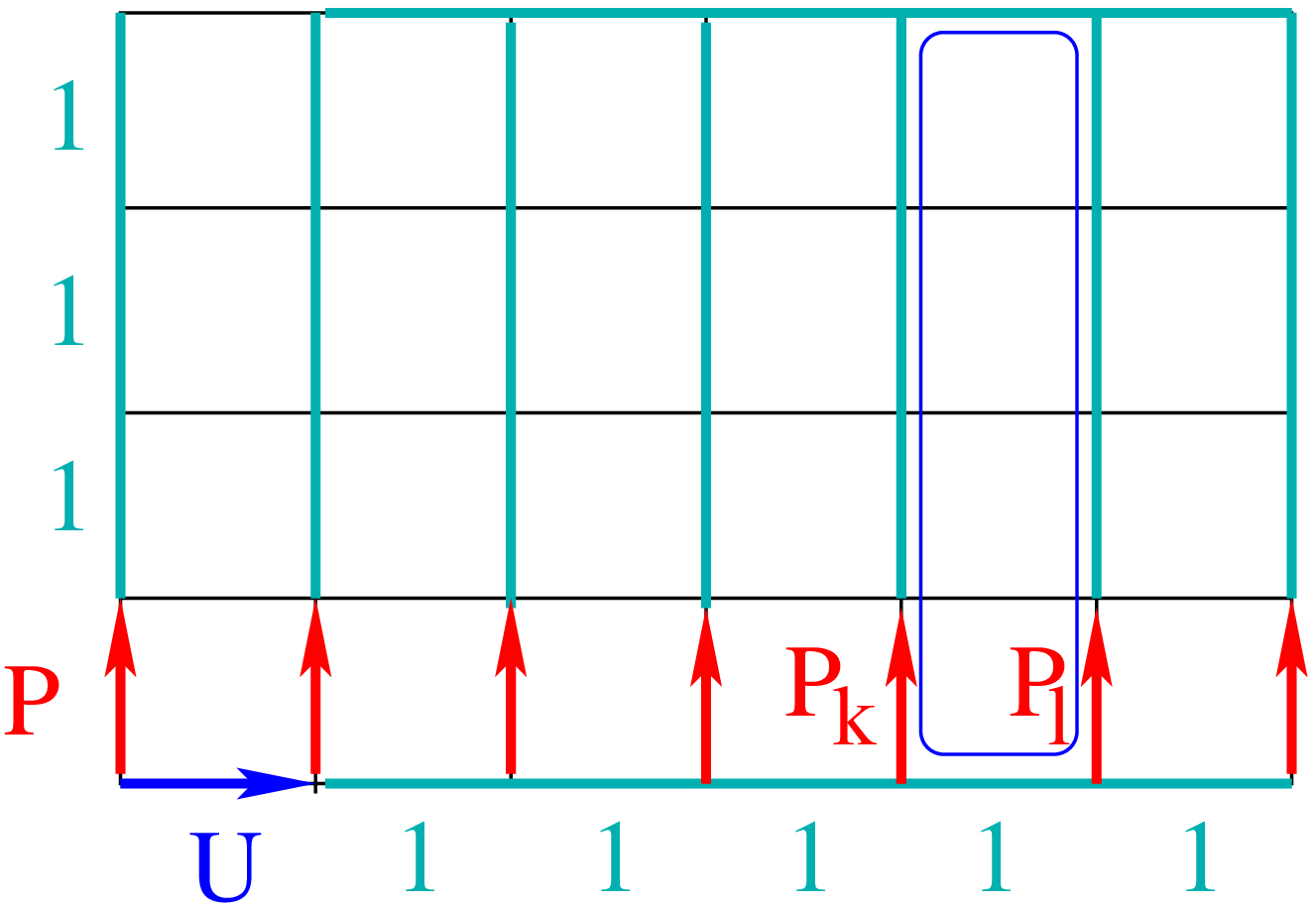} \hspace{0.5cm}
  \includegraphics[width=7cm]{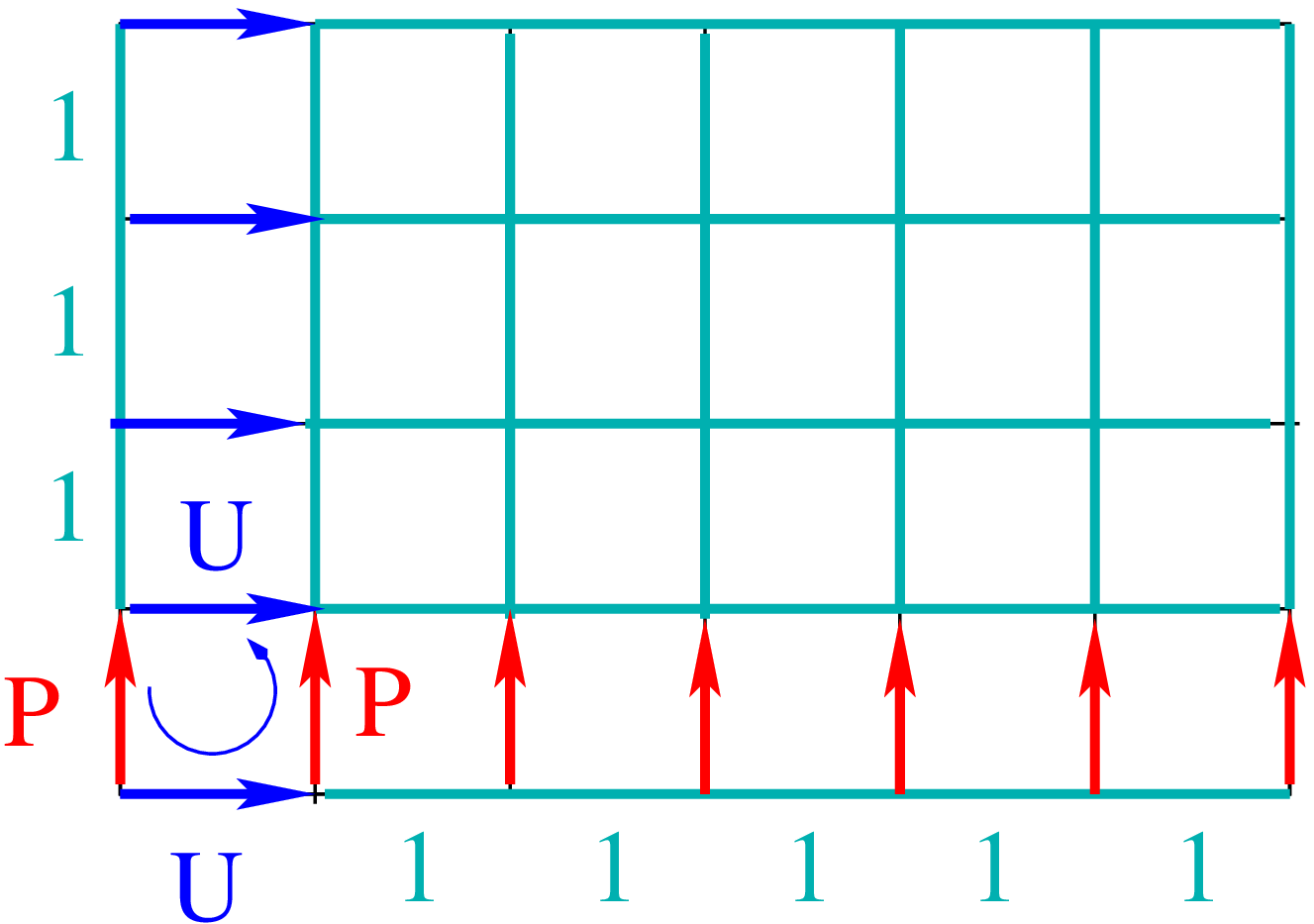}
\caption{\label{fig:2} Gauge fixing step 2 (right) and 
(almost) completely gauge fixed configuration (left). 
}
\end{figure}
Genuinely different vacua states are obtained by identifying 
all states which are related by gauge transformations. 
In order to calculate those, a complete unique gauge fixing is in order.
In a first step, we adopt a Polyakov type of gauge fixing: 
all time like links $U_0(t,\mbx)$ for $t>1$ are gauged to the 
unit element (see figure~\ref{fig:1}, left panel). The only non-trivial
time-like links are those at $t=1$, which thus equal the Polyakov line. 
The gauge transformation $\Omega _1$, $\Omega _2$, etc, are not used 
for this first gauge fixing step. These gauge transformations are 
now employed to gauge almost all spatial links at time-slice $t=1$ 
to the unit element (see figure~\ref{fig:2}, right panel). 
The empty vacuum condition then implies that all spatial links
without arrows in figure~\ref{fig:2} are transformed to the 
unit element. The spatial links $U$ which are far most to the left 
then equal the spatial Polyakov line in this direction. 
We then again use the vacuum condition for the contractible loop 
in figure~\ref{fig:2}, left panel and conclude that 
$$ 
\cP(\mbx_k) \, \one \, \cP^\dagger (\mbx_\ell) \, \one = \one \hbo \Rightarrow \hbo 
\cP(\mbx_k) = \cP(\mbx_\ell)=\cP\, . 
$$
In this gauge and for an empty vacuum state, not only the {\it trace} of the 
Polyakov is constant but also the full Polyakov line including 
off-diagonal parts. After the above gauge fixing, only gauge 
transformation $\Omega _1$ remains unfixed and acts as a residual 
{\it global} colour symmetry. In a final step, we consider 
the plaquette in the bottom left corner of the lattice in
figure~\ref{fig:2}, right panel. Generically, the eigenvalues of the 
spatial and time-like Polyakov lines are different. In order to 
satisfy the vacuum condition for this case, the matrices $\cP$ and $U$ 
are drawn from the Cartan subgroup of $SU(N_c)$ implying: 
$$ 
\cP U \cP^\dagger U^\dagger = \one \hbo \Rightarrow \hbo  [\cP,U]=0 \; . 
$$
Let the spatial Polyakov line $U$ now be an element of the Cartan 
subgroup with all eigenvalues different. The Polyakov line $\cP$ must then be 
an element of the Cartan subgroup in order to describe an empty vacuum:
$$
\cP \; \in \; [U(1)]^{N_c-1} \; . 
$$
For two choices of $\cP$ with at least one eigenvalue 
being different, we obtain two {\it gauge in-equivalent} empty vacua states. 
Note that the Polyakov line homogeneously transforms under gauge 
transformations, 
$$ 
\cP^\Omega (\mbx) \; = \; \Omega (\mbx)  \cP(\mbx) \Omega^\dagger
 (\mbx) \; , 
$$
implying that its eigenvalues are gauge invariant. 
Each $U(1)$ is spanned by a compact angle variable $\alpha _n$, $n=1 \ldots 
N_c-1$, which span the space of global minima of the classical 
Yang-Mills action - the moduli space. For each direction, the 
corresponding Polyakov line can be chosen from the $ [U(1)]^{N_c-1}$ 
subgroup. Hence, the moduli space is at least spanned by the 
$$ 
 [U(1)]^{d(N_c-1)}
$$
group manifold. Note that if, for a particular direction, the eigenvalues 
of the Polyakov line are $f$ fold degenerate, the Polyakov lines 
of the other direction can be chosen from a $SU(f)$ subgroup and still 
satisfy the empty vacuum condition. Hence, the moduli space is slightly 
larger than the space spanned by the $U(1)$ groups only. 

\vskip 0.3cm 
We finally make two comments: 

\medskip 
(i) We point out that the {\it trace} of the Polyakov line 
being different is a {\it sufficient}, but not a {\it necessary }
condition for two states being at different points in moduli space: 
assume that two Polyakov lines possess different eigenvalues but 
the same sum of all eigenvalues. They would belong to different 
points in moduli space, but their trace would be equal. 

\medskip 
(ii) A perturbative treatment should involve a summation over all 
states with minimal (global) action. This implies an integration 
over the moduli space. Note that the standard perturbation theory 
merely chooses one state (i.e., the state of vanishing gauge potential 
or the state of all unit links in lattice formulation) of the moduli space. 
The integration over the moduli space would most likely remove the 
colour states from the theory thus inducing confinement. It would not, 
however, provide a confinement scale. Perturbation theory with 
an integration over the moduli space as well as the phenomenological 
implications of the moduli space integrations in effective quark 
theories will be explored elsewhere. 

\subsection{Yang-Mills quantum vacuum }

The flat directions of the ``empty'' vacuum, discussed in detail in 
the previous subsection, are lifted if quantum fluctuations 
are considered. The symmetry of the quantum effective action 
collapses to the discrete centre symmetry. Introducing 
the centre elements of $SU(N_c)$ by 
\be 
z_m \; = \; \exp \Bigl\{ \I \, \frac{2 \pi }{N_c} \, m \Bigr\} , \hbo 
m = 1 \ldots N_c , 
\label{eq:qv1}
\en 
the centre transformed lattice configuration $\{U^c\}$ is obtained 
by multiplying all time-like links at a given time slice $t_0$ by 
$z_m$: 
\bea 
U^c _0 (t_0,\mbx) &=&  z_m \, U _0 (t_0,\mbx) \; , \hbo 
\hbox{for} \; \; \; \forall \mbx \; , 
\label{eq:qv2} \\
U^c _\mu (t,\mbx) &=&  U _\mu (t,\mbx) \; , \hbo \; \; \; \; \; \;
\hbox{else} \; . 
\label{eq:qv3} 
\ena 
If the lattice action in pure Yang-Mills theory consists of a collection of 
contractible loops (such as the Wilson action which is constructed 
from the smallest of such loops - the plaquette), 
the action is invariant under the above centre transformation. 
On the other hand, the Polyakov line (\ref{eq:pv4}) transform 
homogeneously: 
\be 
\cP[U^c](\mbx) \; = \; z_m \, \cP[U](\mbx) \; . 
\label{eq:qv4} 
\en
Any ergodic Monte-Carlo simulation on a finite lattice 
necessarily averages over the centre copies in a democratic way 
implying that 
$$ 
\left\langle P[U] \right\rangle \; = \; \left\langle P[U^c] \right\rangle 
\; = \; z_m \; \left\langle P[U] \right\rangle 
\hbo \Rightarrow \hbo \left\langle P[U] \right\rangle = 0 \; , 
$$
which is also true in the high temperature phase of Yang-Mills theory. 
Showing that centre sector transitions imply confinement hence 
request more subtle arguments involving centre invariant expectation values. 

\vskip 0.3cm 
To this aim, let us directly consider the static quark antiquark 
potential $V(r)$ as inferred from the Polyakov line correlation 
function (\ref{eq:pv6}). Instead of the Polyakov line expectation 
value, we consider its spatial average in relation to a reference 
$P(0)$ on the lattice. We observe
$$ 
\Bigl\langle P(0) \, \sum _{\mbx} P(\mbx) \Bigr\rangle \; \propto \; 
\sum  _{\mbx}  \mathrm{e}^{-V(r)/T } \; = \; \hbox{finite} 
\hbo \Rightarrow \hbo \lim _{r \to \infty} \, V(r) \to \infty \; , 
$$
and, hence, confinement. If on the other hand the potential 
approaches a finite value for $r \to \infty$, 
the correlator necessarily behaves as 
\be 
\lim _{r \to \infty} \bigl\langle P(0) \, P(\mbx) \bigr\rangle 
\; = \; \hbox{finite}  \hbox to 2 true cm {\hfill or \hfill } 
\Bigl\langle P(0) \, \sum _{\mbx} P(\mbx) \Bigr\rangle  
\to \infty \; . 
\label{eq:qv4b} 
\en
In view of (\ref{eq:qv4}), the latter equation could mean that 
centre sector disorder does not occur at length scales set 
by the lattice size. In the next subsection, we will show that 
the latter condition is in line with what is usually referred to 
as the spontaneous breakdown of centre symmetry. This 
breakdown occurs at high temperatures leaving us with 
the so-called quark-gluon plasma phase.

\subsection{Spontaneous breaking of centre symmetry }

\begin{figure}
  \includegraphics[width=7.5cm]{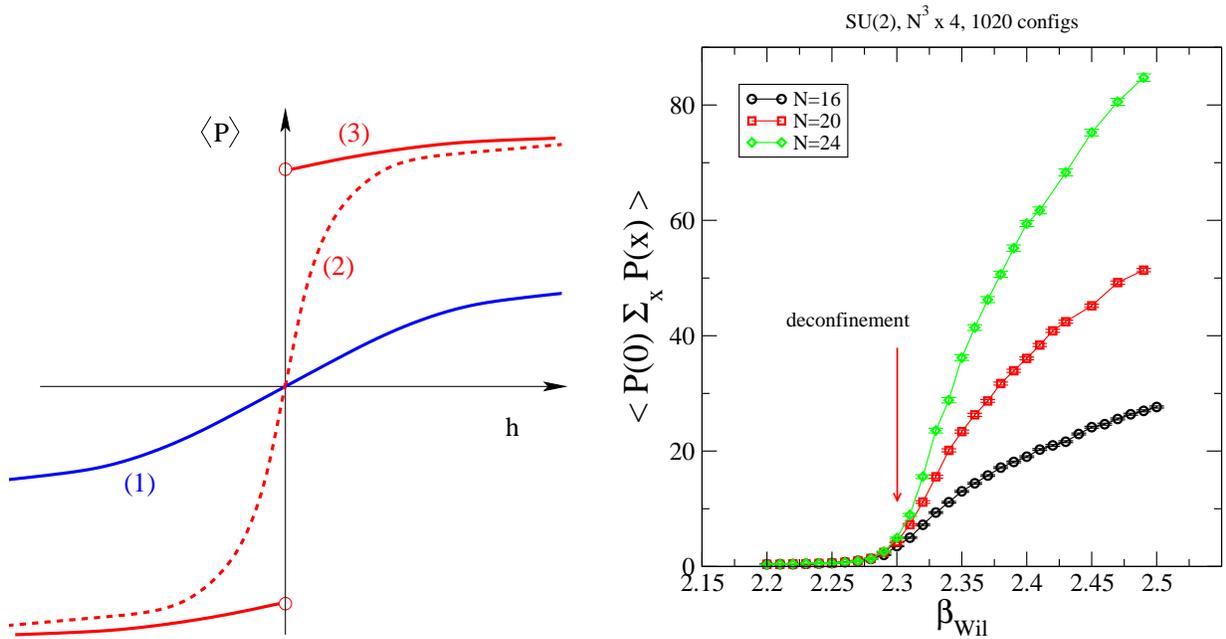}  \hspace{0.5cm}
  \includegraphics[width=8cm]{pol_sq.eps} 
\caption{\label{fig:3} Sketch of the response of the Polyakov line 
expectation value to external centre symmetry breaking (left). 
The response function of the Polyakov line to an external field 
as function of $\beta _\mathrm{Wil}$ for several spatial volumes (right). 
}
\end{figure}
One way to reveal the spontaneous breaking of centre symmetry 
using ergodic Monte-Carlo simulations is to add a centre symmetry 
breaking source term to the action and controlling the strength of this 
term by a parameter, let us say, $h$, which is reminiscent of 
a magnetic field in an Ising type setting. 
To be explicit, 
we will study SU(2) lattice gauge theory using a Euclidean space-time lattice 
represented by a $N_s^3 \times N_t$ grid with lattice spacing 
$a$. Using the Wilson action, the partition function is given 
\bea 
Z _\mathrm{YM} &=& \int {\cal D}U \;  \exp \{ S_\mathrm{Wil} + 
S_\mathrm{break} \} \; , 
\label{eq:qv5}  \\ 
S_\mathrm{Wil} &=& \frac{\beta_\mathrm{Wil}}{N_c} \, \sum _{x, \mu<\nu} \; 
 \tr \left\{ 
U_\mu(x) \; U_\nu(x+\mu) \; U^\dagger _\mu (x+\nu) \; 
U^\dagger _\nu (x) \; \right\} \; , 
\nonumber \\ 
S_\mathrm{break} &=& h \, \sum _{\mbx} \; P(\mbx) \; , 
\nonumber 
\ena
where we have used the Polyakov line (\ref{eq:pv4}) to test the response 
of the theory to explicit centre symmetry breaking.

\vskip 0.3cm 
In the confinement phase, the Polyakov line expectation value 
$\langle P \rangle $  
linearly responds to the presence of an external centre symmetry breaking. 
This is illustrated in figure~\ref{fig:3}, left panel, curve (1). 
Let us now consider the quark gluon plasma, i.e., the deconfinement phase. 
For finite lattice volume, the Polyakov line expectation value necessarily 
vanishes at $h=0$ and is otherwise a smooth function of $h$. 
This is illustrated by curve (2) in  figure~\ref{fig:3}, left panel. 
Only in the infinite volume limit, the $\langle P \rangle $ becomes 
a discontinuous function which approaches a finite value in the limit 
$h \to 0$ (see curve (3) in  figure~\ref{fig:3}, left panel). 

\vskip 0.3cm 
How can we anticipate the spontaneous breaking of the centre symmetry 
by means of a finite volume ergodic Monte-Carlo simulation? 

A straightforward way is to study the gradient of $\langle P \rangle (h)$ 
at $h=0$. This gradient should rise beyond bounds with increasing 
volume. Hence, we find 
\bea 
\frac{ d \langle P \rangle (h) }{dh } \Big\vert _{h=0} &=&
\lim _{h \to 0} \left[ \Bigl\langle P(0) \sum _{\mbx} P(\mbx) 
\Bigr\rangle \; - \; \Bigl\langle P(0) \Bigr\rangle \; 
\Bigl\langle \sum _{\mbx} P(\mbx) 
\Bigr\rangle 
\right] 
\nonumber \\ 
&=& \Bigl\langle P(0) \sum _{\mbx} P(\mbx) \Bigr\rangle 
 \Big\vert _{h=0} \; \to \; \infty \; . 
\label{eq:qv5b}
\ena 
This agrees with the idea of ceasing centre sector transitions 
in the deconfinement phase (see~(\ref{eq:qv4b})). 
Figure~\ref{fig:3}, right panel, shows the 
response function (\ref{eq:qv5b}) for a $N^3 \times 4$ lattice 
as a function of $\beta _\mathrm{Wil}$ for several values of $N$. 
For this lattice geometry, the deconfinement phase is  
attained for $\beta \ge \beta_\mathrm{dec} \approx 2.3$. 
We observe that the response function is basically independent 
of the spatial size in the confinement phase for $\beta < \beta_\mathrm{dec}$ 
while it roughly scales with the volume in the gluonic plasma phase.

\section{Fermi-Einstein condensation in the Schwinger model \label{sec:Schw} } 

It was already observed in~\cite{Langfeld:2009cd} that 
centre sector transitions have far reaching phenomenological consequences 
for gauge theories with dynamical fermions at finite densities (such as QCD): 
for an even number of colours, the centre transformed background field 
can be viewed as imposing {\it periodic boundary conditions} for the 
matter fields. This opens the possibility that the centre dressed 
fermions undergo condensation similar to that which is known as Bose-Einstein
condensation although the fermions are described by anti-commuting 
Grassmann fields. In analogy, this new phenomenon has been called 
{\it Fermi-Einstein condensation}~\cite{Langfeldect2010}. 
For this scenario, centre transitions are essential hence relegating 
Fermi-Einstein condensation to the confining phase at intermediate values 
of the chemical potential. Consequently, Fermi-Einstein condensation 
evades the spin-statistic theorem since the fermion fields 
cannot be viewed as asymptotic states. 

\vskip 0.3cm 
Here and in the next section, we further explore Fermi-Einstein condensation: 
(i) the U(1) gauge theory with fermions at finite densities, i.e., the 
Schwinger model, bears all prerequisites for Fermi-Einstein condensation. 
Because it can be solved analytically, it is also the ideal testbed 
to work out the consequences of the condensation. 
(ii) The rise of Fermi-Einstein condensation is QCD-like theories 
with an odd numbers of colours is less clear cut. We therefore investigate 
a QCD inspired  SU(3) effective quark model to explore its possible existence 
and phenomenological consequences. 

\subsection{Schwinger-model essentials}

Here we illustrate the role of the centre with the finite temperature
Schwinger model in a spatial box of length $L$. The fermion field is 
anti-periodic and the gauge potential periodic in Euclidean time with period 
$\beta=1/T$. Only configurations with vanishing instanton number
contribute to thermodynamic potentials such that we may assume that the fields 
are periodic in the spatial direction.  Thus we
consider 
the Schwinger model on the Euclidean torus $[0,\beta]\times [0,L]$
with volume $V=\beta L$. Only at the end of the calculation do we let the
spatial extension $L$ tend to infinity. 

To continue we decompose the gauge potential as \cite{Sachs:1992pa}
\be
A_0=\frac{2\pi}{\beta}h_0+\partial_0\lambda-\partial_1\phi,\quad
A_1=\frac{2\pi}{L}h_1+\partial_1\lambda+\partial_0\phi
\label{schwinger1}
\en
with constant toron fields $h_0$ and $h_1$. In the zero-instanton sector
the mapping $\{h_\mu,\phi,\lambda\}\to A_\mu$ is one-to-one if we assume
that the periodic functions $\lambda$ and $\phi$ integrate to zero.
Adding an integer either to $h_0$ or to $h_1$ is equivalent to performing a
gauge transformation with winding and thus we must further assume that
$0\leq h_\mu<1$. 
The Polyakov-line is given by 
\be 
P(x_1) \; = \; \exp \left\{ \I \int _0^\beta A_0 \, dx_0 \right\} \; = \; 
\exp \{ 2\pi\I \; h_0 \} \; \exp \left\{ - \I \partial _1 \int _0^\beta 
\phi (x) \, dx_0 \right\} \; . 
\label{schwinger2a}
\en
Using the $[0,1]$ periodicity of the partition function in $h_0$, the shift
\be 
h_0 \; \to \; h_0 \; + \; \alpha 
\label{schwinger2b}
\en
transforms the Polyakov line expectation value as 
\be 
\langle P \rangle \; \to \exp \{ 2\pi\I \; \alpha \} \, \langle P
\rangle \; , \hbo  
\alpha \in [0,1] 
\label{schwinger2c}
\en
and is identified as $U(1)$ centre transformation of the gauge
fields. Hence, integration over the toron field $h_0$ implies an average 
over the centre sectors of the theory. 

\vskip 0.3cm 
The Jacobian of the transformation (\ref{schwinger1}) can
be calculated by expanding all fields in Fourier modes and this way one finds
\cite{Sachs:1992pa} 
\be
{\mathcal D}A_\mu=(2\pi)^2{\det}'(-\triangle){\mathcal D}\phi{\mathcal D}\lambda\,dh_0dh_1,
\label{schinger3}
\en
where 
the primed determinant is the product of all \emph{positive eigenvalues} of the operator $-\triangle$
on the torus. The functional integral over the
gauge functions $\lambda$ cancels out in expectation values
for gauge invariant objects and the $\phi$-dependence of the fermionic 
determinant can be calculated explictly by integrating the chiral 
anomaly~\cite{Blau:1988jp}. The resulting functional 
integral over the periodic function $\phi$ (which must integrate to
zero) is Gaussian and leads to the following expression for the grand
canonical partition  function 
$Z(\beta,L,\mu) = \hbox{tr} \left(\E^{-\beta H+\mu N}\right)$:
\bea
Z(\beta,L,\mu) &=& 
(2\pi)^2
\sqrt{\frac{{\det}'(-\triangle)}{{\det}'(-\triangle+m_\gamma^2)}} 
\int_0^{1} dh_0 \; z(h_0; \, \beta,L,\mu)
\label{schwinger3} \\ 
z(h_0; \, \beta,L,\mu) &=& \int _0^1 dh _1 \; \det(\I\fdirac_{h,\mu})\,,
\label{schwinger3b}
\ena 
where $m_\gamma=e/\sqrt{\pi}$ is the induced photon mass, and 
$z(h_0; \, \beta,L,\mu)$ is the {\it centre sector partition function} 
specified by a particular value of $h_0$. 
The Dirac operator is given by 
\be 
\I\fdirac_{h,\mu} \; = \; \I\fdirac \, - \, 
\left( \frac{2\pi}{\beta } \, h_0 - \I\mu \right) \, \gamma^0 
\, - \, \left(\frac{2\pi }{L} h_1 \right) \, \gamma^1  \; . 
\label{eq:k1}
\en 
The fermionic determinant $\det(\I\fdirac_{h,\mu})$ is an elliptic
function of the constant toron fields $h_0,h_1$ and depends
on the chemical potential. 
We use anti-periodic boundary conditions in time direction 
and periodic ones in spatial direction.

\subsection{Centre-sector partition function} 

Let us further study the grand-canonical partition function 
$z(h_0; \, \beta,L,\mu)$ for a given centre sector. 
This study emulates the spontaneous breaking of centre symmetry where 
the theory is subjected to a ``frozen'' value of $h_0$. We stress that 
such the spontaneous breakdown cannot occur in 2 dimensions due to 
severe infra-red divergences induced by the would-be Goldstone bosons. 
We will however find that the suppression of centre sector transitions 
will lead to the wrong physics, and we believe that this is a generic 
truth valid also in higher dimensions. 

\vskip 0.3cm 
To recover the familiar Fermi sphere physics, it is sufficient 
to consider large spatial volumes for the moment. 
In this limit, we will be able to analytically perform the integration over 
the spatial component $h_1$. To start with, we find: 
\be
\det(\I\fdirac_{h,\mu})  \; = \; 
\E^{-\beta L f(\mu,\beta,L,h_0)} \; \E^{-\beta \,  E_{\rm Cas}(L,h_1) } 
\label{eq:kk2}
\en
where $E_{\rm Cas}(L,h_1) $ is the $\mu $ independent 
Casimir energy (details of the calculation are left to 
appendix~\ref{app:schwinger}): 
\be 
E_{\rm  Cas} \; = \; -\frac{\pi}{6L}+\frac{\pi}{2L}\left(1-2h_1\right)^2.
\label{eq:kk3}
\en
The $\mu $ and temperature dependence is encoded in the so-called 
'free energy density'
\be 
f \; = \; - \frac{1}{L \beta} \sum _{n\in\Z} 
\ln  \Bigl[ \left(1+\E^{2\pi\I h_0} \E^{-\beta (E_n + \mu)}\right)
\left(1+\E^{-2\pi \I h_0} \E^{-\beta(E_n - \mu)} \right) \Bigr] \; , 
\label{eq:kk4}
\en 
where $L \, E_n = 2\pi \vert n - h_1 \vert $. 
For $L/\beta\to\infty$ the free energy density does not depend on the
constant gauge field $h_1$. 
In a theory with {\it fixed} centre sector, i.e., with a {\it frozen value} 
$h_0$, the baryon number density would be given by
\be
\rho_B=\frac{1}{L}\langle Q\rangle=-\frac{\partial f}{\partial\mu}
\stackrel{L\to\infty}{\longrightarrow}
\frac{1}{\pi}\int_0^\infty
dp\left\{
\frac{1}{\E^{2\pi\I h_0} \E^{\beta (p-\mu)}+1}-
\frac{1}{\E^{-2\pi\I h_0} \E^{\beta (p+\mu)}+1}\right\}\,.
\label{eq:kk5}
\en
For the trivial centre sector $h_0=0$, we obtain the standard result 
of a free Fermi gas forming a Fermi sphere at finite values for $\mu $. 
On the other hand, for the sector $h_0=1/2$, we obtain a scenario as if 
the fermions were to acquire Bose-statistics: 
$$ 
f(\mu,h_0=1/2)\stackrel{L\to\infty}{\longrightarrow}
-\frac{1}{\pi\beta }\int_0^\infty \!dp\; \Big\{\log\left(1- \E^{-\beta
  (p-\mu)}\right) + \log\left(1-\E^{-\beta (p+\mu) }\right)\Big\} \, , 
$$
Note that the latter free energy logarithmically diverges when 
$p \to \mu $. This is the usual instability due to condensation and 
corresponds to Fermi-Einstein condensation in the present context. 
Note that this singularity is integrable upon the integration over 
the centre sectors $h_0$. 

\vskip 0.3cm 
In the remainder of this subsection, we will show that 
the assumption of a fixed centre sector will lead to the wrong 
physics in the case of the Schwinger model. 
The key observation is that the only excitations of the model 
are fermion anti-fermion bound states. The theory lacks any 
physical states which would couple to the chemical potential and 
would give rise to non-vanishing fermion density. 
The baryon density ought to vanish even in the case of non-vanishing 
values of the chemical potential. 
An inspection of (\ref{eq:kk5}) shows that this is not the case 
if we consider a fixed value of $h_0$ only. 
The momentum integral can be performed leaving us with ($T=1/\beta$): 
\be 
\pi \, \rho_B \; = \; \mu \; + \; T 
\; \Bigl[ \ln \left( 1 + \E^{-2\pi \, \I \, h_0} \, \E^{\mu/T} \right) 
\; - \; \ln \left( 1 + \E^{2\pi \, \I \, h_0} \, \E^{\mu/T} \right) 
\Bigr] \; , \hbo h_0 \not= 1/2 \; . 
\label{eq:kk7}
\en
At $\mu =0$, the contributions from fermions and anti-fermions to 
the baryon density cancels since there is no fermionic mass gap 
in the theory. For $h_0=1/2$, the theory produces a condensed states, 
and the result (\ref{eq:kk7}) cannot be extended to cover this case. 
For the trivial centre sector $h_0=0$, 
the baryon density linearly rises with the chemical potential. 
This is the well known Fermi-sphere behaviour for the 2-dimensional case. 
For any other value of $h_0$, the baryon density is complex which 
renders the findings difficult to interpret in physical terms. 

\vskip 0.3cm 
The dependence of the baryon density on the chemical potential 
for a fixed centre sector is clearly spurious since the 
spectrum of the theory is free from states which carry baryon charge. 
A similar spurious dependence has been encounter in other theories 
as well which have been treated in an approximative way, and has 
been called the {\it silver blaze problem}~\cite{Cohen:2003kd}.

\subsection{Centre sector average }

Considering individually the contributions of the centre sectors (specified by
$h_0$) the free energy density has led to a singularity 
for $h_0=1/2$ associated with the condensation of fermions 
by virtue of periodic boundary conditions in this specific sector. 
In the Schwinger model, we are now in the comfortable situation that 
we can explore the physics of this Fermi-Einstein condensation since 
the centre sector average can be performed analytically. 
To this aim, we point out that the average of the fermion determinant 
over the spatial toron field can be written as (see appendix~\ref{app:theta} 
for details): 
\be
\int_0^1 dh_1 \, \det(\I\fdirac_{h,\mu}) \; = \;
\frac{1}{\sqrt{2\tau}\vert\eta(\I\tau)\vert^2} \sum_p \E^{-\ha \pi\tau p^2+2\pi
  \I p \gamma} \;, \hbo \gamma=h_0+\frac{\I\beta}{2\pi}\mu \; , 
\label{eq:kk10} 
\en
where $\eta(\I\tau)$ is the Dedekind eta-function (see~(\ref{app7})). 
The centre sector average, i.e., the integration over $h_0$, 
can be easily performed and restricts the Matsubara momentum sum 
in (\ref{eq:kk10}) to the trivial momentum $p=0$: 
\be
\int_0^1 dh_0 dh_1 \; \det(\I\fdirac_{h,\mu}) \; = \;
\sqrt{\frac{1}{2\tau}}\frac{1}{\vert\eta(\I\tau)\vert^2} \; . 
\label{eq:kk11}
\en
The Roberghe-Weiss symmetry for a model with global $U(1)$ centre-symmetry 
implies that the partition function can not depend on the imaginary
part of $\mu$. Analyticity suggests that this might also hold for real 
values of the chemical potential. Indeed for the present case, we find 
that any dependence on $\mu $ disappears after the integration over $h_0$. 

\vskip 0.3cm 
From~\cite{AlvarezGaume:1986es} we take the result 
\be
{\det}'^{1/2}(-\triangle)=\beta\vert\eta(\I\tau)\vert^2 \; 
\label{eq:kk12}
\en
to write 
\be
\int_0^1 dh_0 dh_1 \; \det(\I\fdirac_{h,\mu}) \; = \; 
\sqrt{\frac{V}{2}}\frac{1}{{\det}'^{1/2}(-\triangle)}.
\label{eq:kk15}
\en 
Inserting the latter finding into the grand canonical partition 
function (\ref{schwinger3}), the square root of the determinant
cancels, and we finally obtain 
\be
Z(\beta,L,\mu) \; = \; \sqrt{\frac{V}{2}}\frac{1}{
\sqrt{{\det}'(-\triangle+m_\gamma^2)}} \; . 
\label{eq:kk16} 
\en
The key observation is that the $\mu $ dependence has disappeared 
implying that the centre sector average has solved the 
{\it silver blaze problem}: the baryon density vanishes independent 
of the value for the chemical potential as it should 
for the Schwinger model. 

\vskip 0.3cm 
It is instructive for QCD model building to investigate how the 
centre sector average eliminates the $\mu $-dependence. To this aim, 
we reconsider from the partition function (\ref{schwinger3}) the factor 
\be
\int _0^1  dh_0 \; \det(\I\fdirac_{h,\mu}) \, = \, \E^{-\beta E_{\rm Cas}(L)} 
\int _0^1  dh_0 \; 
\prod_{n\in\Z} \left(1+\E^{2\pi\I h_0} \E^{-\beta (E_n + \mu)}\right)
\left(1+\E^{-2\pi \I h_0} \E^{-\beta(E_n - \mu)}\right) \; ,
\label{eq:kk17}
\en
where $L \, E_n = 2\pi \vert n - h_1 \vert $. 
Expanding the product yields terms such as 
$$ 
\exp \{2\pi n \, \I   h_0 \} \; \exp \{ - \beta \, n \, \mu \} \; . 
$$
These terms vanish upon the integration over $h_0$ {\it unless } 
we have $n=0$. Hence, the integral (\ref{eq:kk17}) does not depend 
on $\mu $. This finding has a direct physical interpretation: 
$n=0$ only occurs if as many fermions as anti-fermions contribute 
to the product. Hence, the $h_0$ integration eliminates 
all states which carry a net baryon charge thus solving the 
{\it silver blaze problem}.

\subsection{Volume studies } 

\begin{figure}
  \includegraphics[width=8cm]{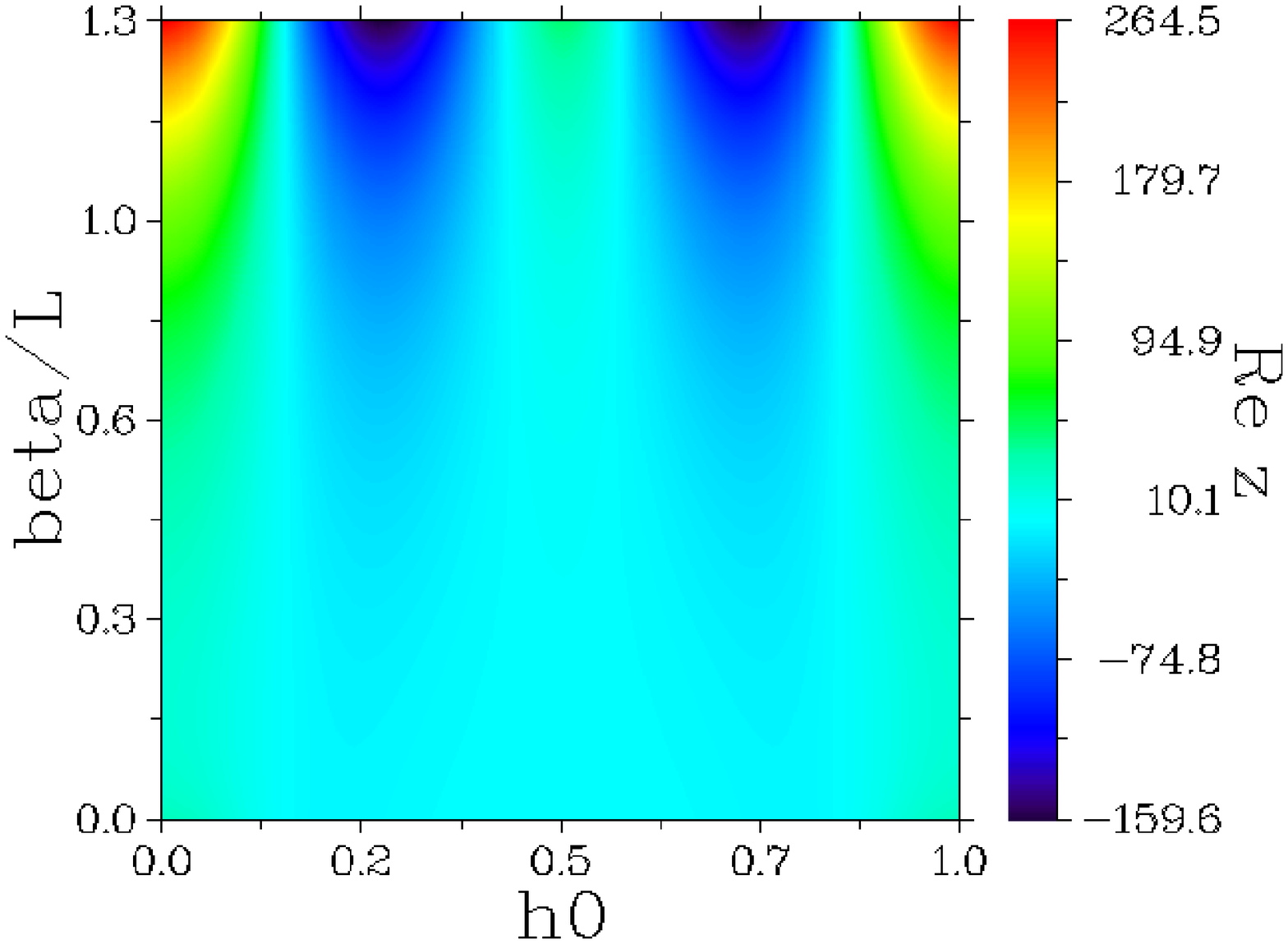}  \hspace{0.5cm}
  \includegraphics[width=8cm]{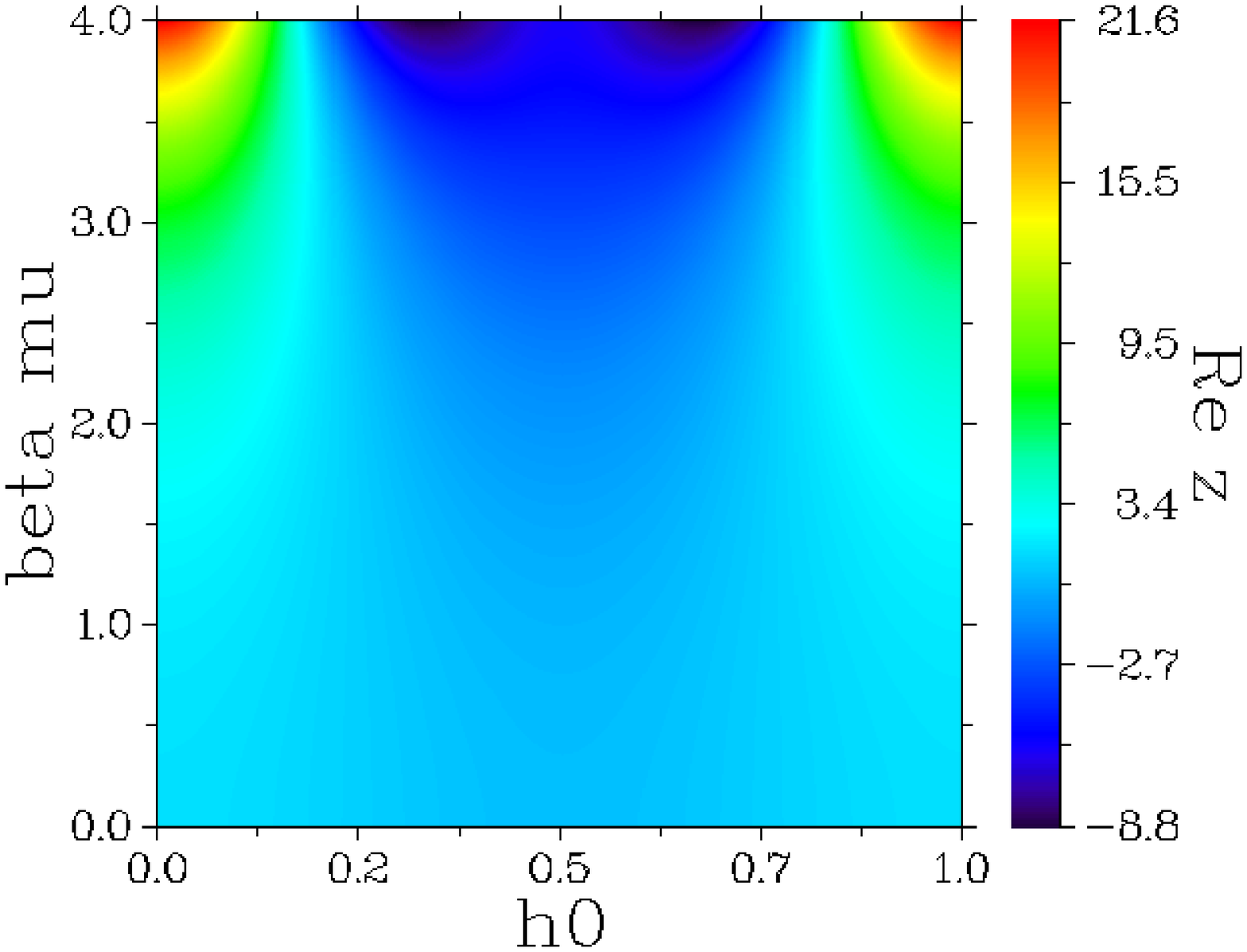}  \hspace{0.5cm}
\caption{\label{fig:vol} The real part of the differential contribution $z$
  in (\ref{schwinger3b}) to the fermion determinant as a function of $h_0$ and 
$\beta/L$ for $\beta \mu = 4$ (left). The real part of $z$ as a function 
of $h_0$ and $\beta \mu $ for $\beta /L=1$ (right). 
}
\end{figure}
The question whether a particular centre sector is singled out or 
whether a more democratic average over the centre sectors is in order 
should be answered by the theory itself. For this investigation, we will here
consider the large volume limits $L \to \infty$ and $\beta \to  
\infty $ and combinations of it. 

\vskip 0.3cm 
In (\ref{schwinger3}), we decomposed the grand canonical partition function 
$Z$ into a part from dynamical fields and a part 
$ z \left(h_0 ; \,\beta/L, \beta \, \mu \right)$ featuring 
the contributions from the toron fields. 
From the studies in the previous subsection, we already know that 
$Z_\mathrm{fer}$ is actually independent of $\beta \, \mu $. 
For a given value of the fugacity $\mu /T = \beta \mu$, we have 
calculated $z$ as a function of $h_0$ and $\beta /L$. 
The colour coded result is shown in figure~\ref{fig:vol}, left panel. The
important  
observation is that the largest contribution arises from the trivial 
centre sector around $h_0=0$ at least for $\beta \mu >1$ 
(note that the $z$ is periodic, i.e., $z(h_0)=z(h_0)+n$, $n$ integer). The
maximum at  $h_0=0$ even diverges in the limit $\beta /L \to \infty$. 

\vskip 0.3cm 
Here is a lesson to learn if it comes to Monte-Carlo simulations. 
Assume that we were to estimate the $h_0$ integral using a Metropolis 
Monte-Carlo method. First of all we note that for $h_0\not=\{0.5, 1\}$ 
the determinant is complex. For small enough $\beta \mu$, let us say 
$\beta \mu < 1$, we could use the real part of the determinant as a
reweighting factor for the simulation. From figure~\ref{fig:vol}, left panel, 
it is already clear that the method samples the region around $h_0=0$ to a large
extent.  For $\beta /L \gg 1$, other contributions from larger values of $h_0$ 
would be extremely rare. If we now stick to fixed aspect ratio $\beta /L$, 
we will find that for large values of $\beta \mu $ the real part of the 
determinant develops negative parts invalidating the Monte-Carlo approach 
altogether (see figure~\ref{fig:vol}, right panel). 
The keypoint, however, is that the integral of $z$ over $h_0$ yields a 
constant only dependent on the aspect ratio (see the previous 
subsections). This means that integrating along the 
horizontal axis in figure~\ref{fig:vol}, right panel, must produce 
the same value for every choice of $\beta \mu $. 
Obviously, sampling the whole range of $h_0$ values in the Monte-Carlo 
simulation is of key importance although difficult to achieve for 
$\beta /L \gg 1$, i.e., in the thermodynamic limit $L \to \infty $ 
with $T\not=0$ fixed. Our findings suggest that the infinite 
volume zero-temperature limit of the QFTs (which we are going to discuss 
below) is delicate: we suggest that these limits  should be taken at fixed
aspect ratio $\tau = \beta /L$ and with one of the variables, $L$ or $\beta $ 
tending to infinity. In this context, the value $\tau =1$ is of
particular relevance for lattice gauge theories, since in this limit the
rotational symmetry is recovered  in the scaling limit of vanishing lattice
spacing.  

\vskip 0.3cm 
\begin{figure}
  \includegraphics[width=8cm]{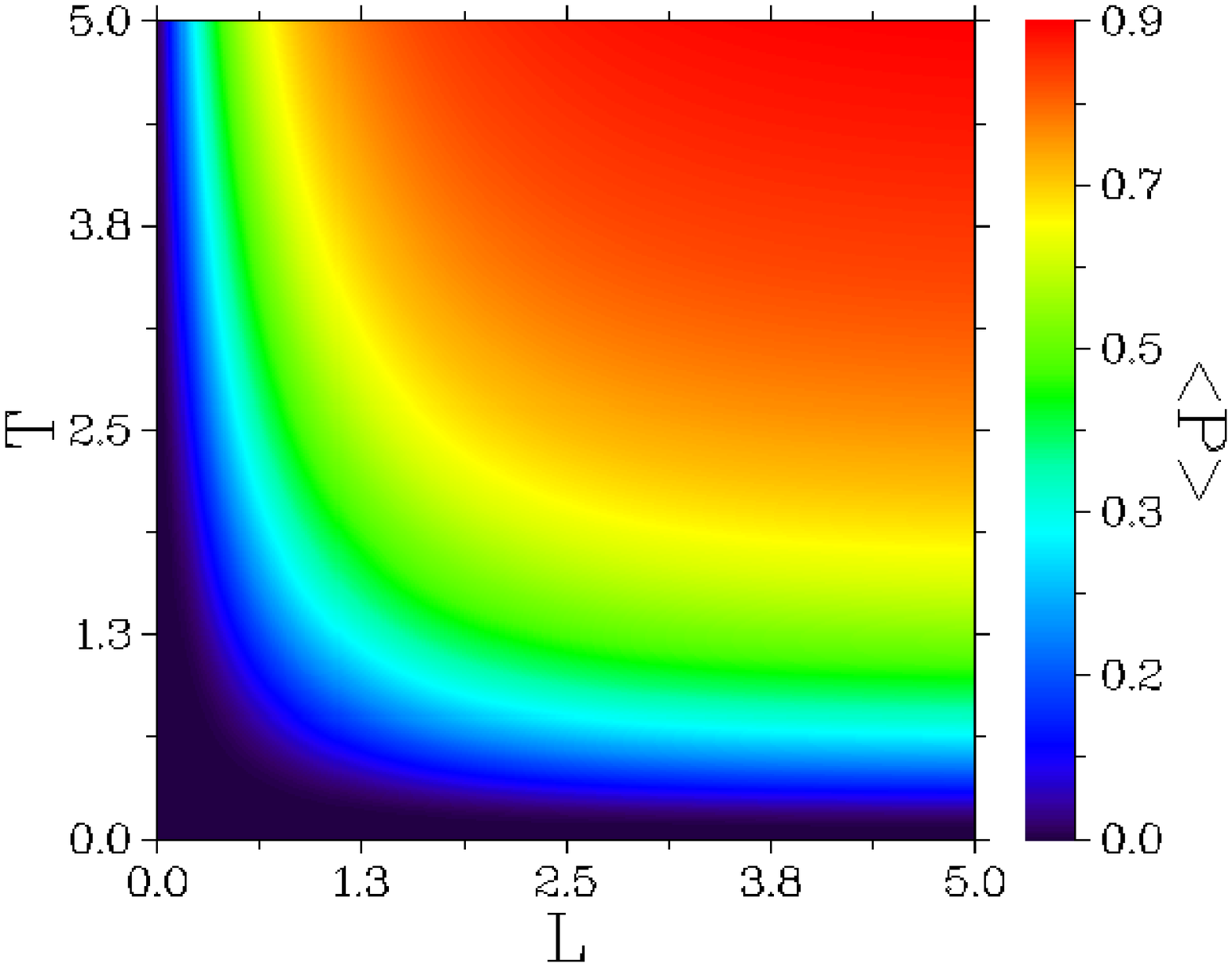}  \hspace{0.5cm}
  \includegraphics[width=8cm]{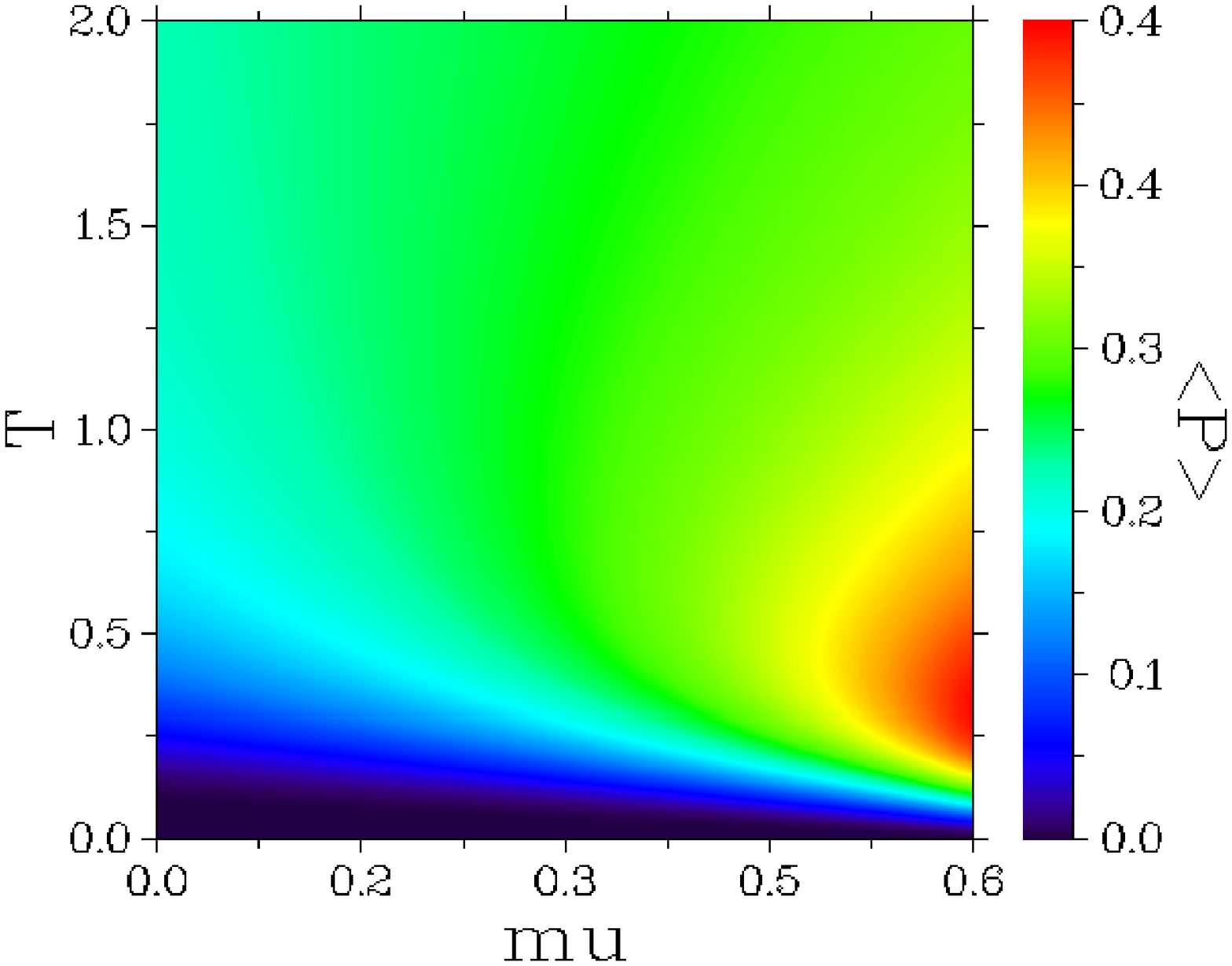}  \hspace{0.5cm}
\caption{\label{fig:s_pol} The Polyakov line expectation value for 
  vanishing chemical potential as a function of the temperature $T$ 
  and spatial size $L$ (left). $\langle P \rangle $ for a fixed 
  aspect ratio $\beta /L=1$ as a function of $\mu $ and $T$ (right). 
}
\end{figure}
Although the centre transitions never cease in the Schwinger model, 
the trivial centre sector contributes an overwhelming part to the 
expectation values. This is also clear from the Polyakov line expectation 
value. This expectation value can be calculated in closed from for arbitrary
lengths  $\beta $, $L$ and at presence of chemical potential $\mu $ 
(details of the calculation can be found in appendix~\ref{app:pol}): 
\be
\left\langle P\right\rangle \; = \; \E^{q\beta \mu} \; 
\exp\left(-\frac{\pi\beta m_\gamma}{4} 
\coth\left(\frac{m_\gamma L}{2}\right)\right) \; , 
\label{eq:kk20}
\en
where $m_\gamma=e/\sqrt{\pi}$ is the dynamical generated photon mass 
which sets the scale of the theory. 

\vskip 0.3cm 
Let us firstly discuss the case of vanishing chemical potential. 
Our findings for this case are summarised in figure~\ref{fig:s_pol}, 
left panel. Note that the Polyakov line expectation value 
$\langle P \rangle $ always vanishes for any fixed spatial size 
$L$ in the zero temperature limit $T \to 0$. 
On the other hand for a fixed temperature, $\langle P \rangle $  
{\it increases} with increasing system size $L$, and potentially 
approaches quite large value in the infinite volume limit: 
$$ 
\lim _{L \to \infty} \left\langle P\right\rangle  \; = \; 
\exp \left\{ - \frac{\pi}{4}  \frac{m _\gamma}{T} \right\} \; . 
$$
We will see below that this behaviour will be mirrored by the 
SU(2) Higgs theory. 

\vskip 0.3cm 
Secondly, we consider the case of a non-vanishing chemical potential. 
For a fixed aspect ratio $\beta/L =1$, figure~\ref{fig:s_pol}, 
right panel, shows the Polyakov line expectation value as a function 
of $\mu $ and $T$. We point out that the fermion determinant is complex 
implying that the expectation value of a phase, in particular the 
Polyakov line, is not necessarily bounded anymore. These imaginary parts 
allow for the cancellations which are essential to solve the Silver Blaze 
problem and, at the same time, bear the potential for 
$\vert \langle P \rangle \vert \gg 1$. In the infinite volume limit, we now 
find 
$$ 
\lim _{L \to \infty} \left\langle P\right\rangle  \; = \; 
\exp \left\{ \frac{\pi}{4} \; \frac{(\mu - m _\gamma)}{T} \right\} \; , 
$$
which is bigger than one if the chemical potential exceeds the photon mass.

\section{Fermi-Einstein condensation in a QCD quark model with confinement 
\label{sec:Qconf} }

\subsection{Model building} 

In order to trace out the phenomenology of centre sector transitions
and Fermi-Einstein condensation in a more QCD-type setting, we 
are here going to investigate an effective SU($N_c=3$) quark theory 
in four dimensions.  
The so-called constituent quarks $q(x)$ satisfy the usual anti-periodic 
boundary conditions in Euclidean time direction (and periodic ones in 
the spatial directions)
\be 
q(x+\beta e_0) \; = \; (-1) \, q(x) \; ,\qquad
q(x+\beta e_k) \; = \; q(x),\qquad k=1,2,3 \; .
\label{eq:eff1}
\en 
The quarks possesses the constituent quark mass $m$ and are assumed
to only interact with an homogeneous temporal gauge field specifying 
the centre sector: 
\be
A_0^{(n)} = 2\pi n \, T \, H ,  \hbo 1 \le n \le N_c \, ,
\label{eq:eff2}
\en
where the generator $H$ is from the Cartan algebra, i.e., 
$H = \mathrm{diag}(1,\ldots, 1, 1-N_c)/N_c$. 
The Polyakov line is in the centre of the group and the
centre sector is labeled by $n$ since the trace of the Polyakov line 
is given by 
\be 
P \; = \; \frac{1}{N_c} \, \tr \, \exp \left\{ \I \int _0^\beta dx_0 \; 
A_0^{(n)} \right\} \; = \; z_n, \hbo 
z_n \; = \; \exp \left\{\frac{2\pi\I}{N_c} \, n \, \right\} \; . 
\label{eq:eff3}
\en
The crucial observation~\cite{Langfeld:2009cd} is that by means 
of a Roberghe-Weiss transformation quarks which are subjected to the 
background field $A_0^{(n)}$ can be considered as quarks manoeuvring 
in the trivial background $A_0^{(N_c)}$ but with changed boundary conditions: 
\be 
A_0^{(n)}, \; q(x+\beta e_0) =  - q(x) \hbo \leftrightarrow \hbo 
A_0^{(N_c)}, \; q(x+\beta e_0) =  -z_n q(x) \; . 
\label{eq:eff4}
\en
For even $N_c$, there is the element $z_{N_c/2}=-1$, and it is this 
centre sector which is mapped to the trivial sector but with quarks 
now obeying {\it periodic} boundary conditions. Since it is essentially 
the boundary conditions which dictate the behaviour of the thermodynamical 
potentials, centre-dressed quarks follow Bose statistic and undergo 
condensation if the chemical potential approaches the fermionic mass 
gap~\cite{Langfeld:2009cd}. 
Apparently, this scenario relies on the fact that $(-1)$ is an 
element of the centre of the gauge group. This is only the case 
if the number of colours $N_c$ is even. The more interesting, i.e., the 
more QCD type, case is $N_c=3$ and evades the line of arguments. 
In this subsection, we will explore the phenomenology of 
the centre sector transitions in the realm of an effective quark model 
for $N_c=3$. 

\vskip 0.3cm 
The partition function of our model is given by 
\bea
Z_Q &=&  \sum _{n=1}^{N_c} \, \int {\cal D} q{\cal D} \bar{q} \; 
\exp \left\{\bar{q} \big(\I\dslash + (A_0^{(n)} + \I\mu) \gamma^0 +  \I m \big) q \right\} 
\; = \; \sum _{n=1}^{N_c} \E ^{\,\Gamma ^{(n)}} , 
\label{eq:eff5} \\ 
\Gamma ^{(n)} &=& \ln \,  \det \Bigl( \I\dslash + (A_0^{(n)} + \I\mu) \gamma_0 +
im  \Bigr) \; , 
\label{eq:eff6}
\ena  
where $m$ is the quark mass and $\mu $ is quark chemical potential.
The main difference to the model discussed in~\cite{Langfeld:2009cd} 
is that we take into account that the gluonic sector is
centre symmetric and the only centre sector bias arises from 
the quark determinant. Thus, the sum in (\ref{eq:eff5}) democraticly extends 
over all centre sectors without any further bias to the trivial centre sector.

\subsection{Cold but dense matter} 

The calculation of determinant in (\ref{eq:eff5}) can be performed 
following the techniques developed for the Schwinger model. 
In 4 dimensions the determinant is more severe UV-divergent
than in 2 dimensions but the {\it effective } theory only addresses the 
low-energy modes below a certain physical energy scale. In this approach, the cutoff scale 
$\Lambda $ is finite and acquires a physical interpretation.  Here, 
we are only considering temperatures which are small compared 
to this cutoff scale which allows us to effectively set $T/\Lambda \to 0$. 
Using a cutoff function provided by the Schwinger-proper time 
approach (see appendix~\ref{app:swp}), we find: 
\be 
\Gamma ^{(n)} = \Gamma _0(\beta,L,\Lambda) \; + \; \Gamma  ^{(n)}_\mathrm{den}
(\beta, L, \mu) 
\label{eq:eff15} 
\en
with cutoff-dependent and cutoff-independent contributions
\bea
\Gamma _0(\beta,L,\Lambda) &=& \sqrt{\pi} \; \beta  \Lambda \; \sum_p 
\bigl[ 1 \, - \, \mathrm{erf} 
\left( E(p)/\Lambda  \right)  \bigr] \; , 
\label{eq:eff16} \\ 
\Gamma  ^{(n)}_\mathrm{den}(\beta, L, \mu) &=& 2 \, \sum_p  
\Big\{ \ln \left( 1 \;+\; z_n \, \E^{- \beta (E(p) + \mu)} \right) \, + \, 
\ln \left( 1  +  z_n^\ast \, \E^{ - \beta (E(p) - \mu) } \right)\Big\} \; , 
\label{eq:eff17} 
\ena 
where we have introduced the 1-particle energy by 
\be 
E(p) \; = \; \sqrt{ m^2 + p^2} \; , \hbo 
p \; = \; \frac{2 \pi }{L } \, (n_1,n_2,n_3)^T \; , \; \; \; 
n_k: \; \hbox{integer} . 
\label{eq:eff18} 
\en
The overall factor $2$ in the 4-dimensional case (as compared 
to the 2d Schwinger model) arises from the two spin-orientations 
of the quarks. We point out that the cutoff dependent part 
$\Gamma _0$ is independent from the centre sector number $n$ and the 
chemical potential while those parts $\Gamma  ^{(n)}_\mathrm{den} $ which
do depend on $\mu $ and $n$  are UV finite. 

\vskip 0.3cm 
Let us now consider the case of low temperatures (and $\mu > 0$) for which 
we can neglect the anti-quark contributions to $\Gamma  ^{(n)}_\mathrm{den}$:
\be 
\beta \, m \; \gg \; 1 \hbo \Rightarrow \hbo 
\E^{- \beta (E(p) + \mu)} \, \to \, 0 \; . 
\label{eq:eff19} 
\en 
For such low temperatures, even the mesonic type excitations can be 
neglected, and the partition function is approximately given by 
\be 
Z_Q \; \approx \; \E^{\Gamma _0} \; \sum _{n=1}^{N_c} \;  \prod _p \, 
\Bigl[ 1  +  z_n^\ast \, \E^{ - \beta (E(p) - \mu)} \Bigr] ^2 \; . 
\label{eq:eff20} 
\en 
Let us now specialise to $N_c=3$, and expand the brackets. We 
obtain terms such as 
$$ 
 \left[z_n^\ast \right]^\nu \, \E^{ - \nu \, \beta (E(p) - \mu)} \; . 
$$ 
In order to evaluate the sum over the centre sectors, we use 
$$ 
\sum _{n=1}^3 z_n =0 \, , \hbo 
\sum _{n=1}^3 z_n^2 =0 \, , \hbo 
\frac{1}{3} \, \sum _{n=1}^3 z_n^3  = 1 \, . 
$$
Hence, the centre sector sum eliminates all coloured excitations from 
the partition function by projecting onto states with vanishing $N$-ality 
thus making manifest the confinement of colour. 
For $\mu < 6 m$, we obtain the simple result 
\be 
Z_Q \; \approx \;3\; \E^{\Gamma _0} 
\left[ 1  \; +  \; \sum_{p_1, p_2, p_3} 
\exp \Bigl\{ - \beta \left( E(p_1) + E(p_2) + E(p_3)- 3 \mu \right) \Bigr\}
\right] \; , 
\label{eq:eff21} 
\en 
where in the momentum sum only two of the three momentum can be equal. 
Since our model does not include any binding between the quarks, 
the sum $E(p_1) + E(p_2) + E(p_3)$ can be interpreted as 
the energy of the ``baryon'' in the context of the present model. 
For chemical potentials smaller but close to the constituent quark mass $m$, 
the thermal energy density, i.e., 
\be 
E_\mathrm{therm} (T) \; = \; - \frac{ \partial\;  \ln \, Z_Q}{ 
\partial \beta } \; , 
\label{eq:eff22} 
\en 
behaves like 
$$
E_\mathrm{therm} (T) \; \approx \; 3m \; 
\exp \Bigl\{ - \,  \frac{3( m-\mu) }{T} \Bigr\}   \, , \hbo 
\mu \stackrel{<}{_\sim } m  \; . 
$$
The first excitations in the model are baryonic ones. 
This result should be compared with the case of the ``frozen'' 
centre sector (which, in our simple case, is the free quark model): 
$$
E^\mathrm{free}_\mathrm{therm} (T) \; \approx \; m \; 
\exp \Bigl\{ - \,  \frac{ m-\mu }{T} \Bigr\}   \, , \hbo 
\mu \stackrel{<}{_\sim } m  \; . 
$$
Here, the mass gap is provided by the constituent quark mass, and 
the first excitations which are encountered by increasing the temperature 
stating at $T=0$ are quark excitations. As in the 
Schwinger-model, the centre sector sum solves the Silver-Blaze problem.

\subsection{Condensation of centre dressed quarks } 

\begin{figure}
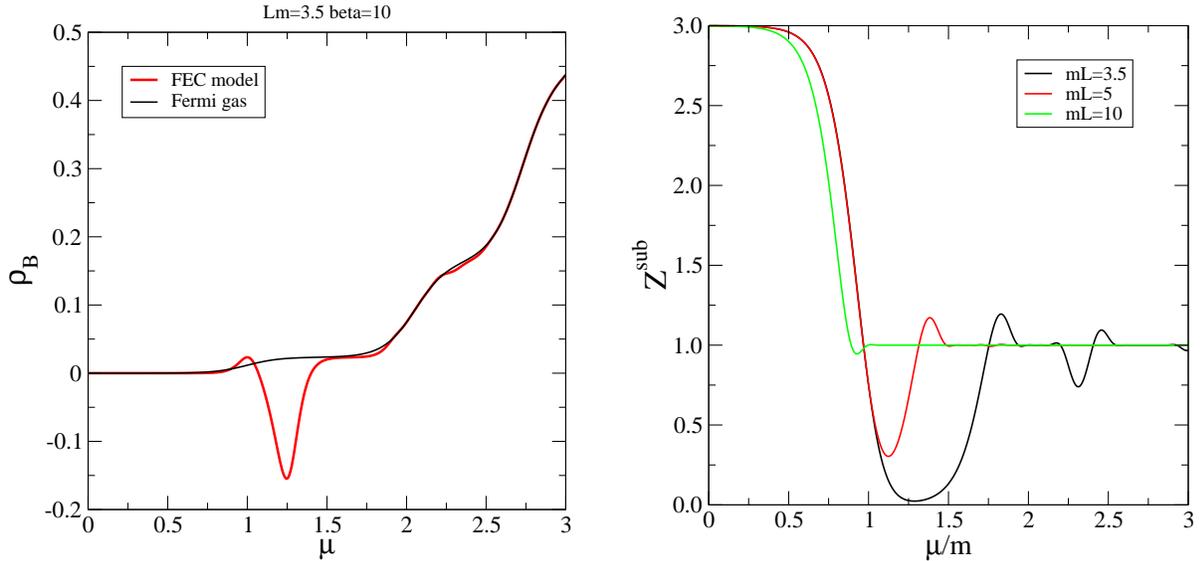

  \includegraphics[width=7.5cm]{rhoB_mu.eps} \hspace{0.5cm}
  \includegraphics[width=7.5cm]{zq.eps}
\caption{\label{fig:cd1} Baryon number as a function of the chemical 
potential for the confining quark model (FEC) in comparison to the standard 
Fermi gas result, left panel. The normalised partition function in
(\ref{eq:eff29}), right panel.
}
\end{figure}
The thermodynamical quantities of the present quark model can also 
be calculated exactly by evaluating e.g.~(\ref{eq:eff17}) by numerical means. 
The constituent quark mass $m$ here sets the fundamental energy scale. 
It turns out that the approximations leading to (\ref{eq:eff21}) are only 
valid for rather low temperatures, i.e., $\beta m \approx 100$. 
In this subsection, we study intermediate temperatures and several 
spatial extensions such as 
\be 
\beta \, m \; = \; 10 \; , \hbo L \, m \; = \; 3 \ldots 10 \; . 
\label{eq:eff23} 
\en 
We are interested in the baryon number $Q_B $ which is accessible by 
\bea 
Q_B &=& T \; \frac{ \partial \, \ln Z_Q}{\partial \mu } 
\; = \; \sum _n \, \omega _n \; \sum _p \, \left[ \frac{ z_n^\ast }{ 
\E^{ \beta (E(p) - \mu)} + z_n^\ast  } \; - \; \frac{ z_n }{ 
\E^{ \beta (E(p) + \mu )} + z_n }  \; \right] \; , 
\label{eq:eff24}  
\ena  
where the centre sector weights are given by 
\be 
\omega _n \; = \; \frac{ \exp \{ \Gamma  ^{(n)}_\mathrm{den} \} }{ 
\sum _n  \exp \{ \Gamma  ^{(n)}_\mathrm{den} \} } \; . 
\label{eq:eff25}  
\en
Note that for $\mu \not=0$ the weights can be complex and therefore
evade a straightforward interpretation as probabilistic weight for 
a given centre sector. An inspection of (\ref{eq:eff17}) shows that 
\be 
\omega _1 \; = \; \omega _2 ^\ast \; , \hbo \omega _3 \, \in \, \R \; , \hbo 
\sum _n  \exp \{ \Gamma  ^{(n)}_\mathrm{den} \}  \, \in \, \R \; . 
\label{eq:eff25b}  
\en
We can always compare our findings with those 
from a free Fermi gas model for which we have 
\be 
\omega _1 \; = \; 0 \; , \hbo \omega _2 \; = \; 0 \; , \hbo 
\omega _3 \; = \; 1 \hbo \hbox{(Fermi gas)} \; . 
\label{eq:eff26}  
\en
The calculation of the centre sector weights is delicate since 
the effective actions $\Gamma _n$ generically are large numbers. 
To facilitate this calculation, we introduce the subtracted actions by 
\be 
\bar{\Gamma} _n  \; = \; \Gamma _n \, - \; \Gamma _\mathrm{max} , \hbo 
\Gamma _\mathrm{max} \; = \; \mathrm{max} \, \Bigl( \hbox{Re} \, \Gamma _n 
\Bigr) \; \; \; \forall n \; . 
\label{eq:eff27}  
\en
It is easy to check that the centre sector weights are actually independent 
of this shift: 
\be 
\omega _n \; = \; \frac{ \exp \{ \Gamma  ^{(n)}_\mathrm{den} \} }{ 
\sum _n  \exp \{ \Gamma  ^{(n)}_\mathrm{den} \} } \; = \; 
 \frac{ \exp \{ \bar{\Gamma}  ^{(n)}_\mathrm{den} \} }{ 
\sum _n  \exp \{ \bar{\Gamma}  ^{(n)}_\mathrm{den} \} } \; . 
 \label{eq:eff28}  
\en
Figure~\ref{fig:cd1}, left panel, shows our finding for the Baryon density 
as a function of the chemical potential $\mu $ for a {\it small } spatial 
volume: 
$$
 m \, L \; = \; 3.5 \; . 
$$
For $\mu/m \approx 1.3 $, we observe that the baryon density strongly 
peaks for the FEC model. This peak is absent in the standard Fermi gas model. 
The reason for this peak are 
cancellations between the centre sector which nullify the partition function. 
To see this, we introduce the normalised partition function by 
\be 
Z^\mathrm{sub} \; = \; \exp \{ \Gamma _n \} \; \E^{- \Gamma _\mathrm{max} } 
\; = \; \sum _n  \exp \{ \bar{\Gamma}  ^{(n)}_\mathrm{den} \} \; . 
\label{eq:eff29}  
\en
Note that $ \Gamma _\mathrm{max} \in \R$ implying that 
a {\it zero} of $Z^\mathrm{sub}$ goes along with a {\it zero} of the full
partition function $Z_Q$ . The quantity $Z^\mathrm{sub} $ also appears in the
denominator of the sector weights  
$\omega _i$. Figure~\ref{fig:cd1}, right panel, shows the normalised 
partition function $Z^\mathrm{sub}$ as a function $\mu $. 
We observe for $m L =3.5$ that $Z^\mathrm{sub}$ almost vanishes 
for the chemical potential corresponding to the peak position in 
the baryon density. Increasing the spatial size, we find 
the near-zero regime of $Z^\mathrm{sub}$ is lifted and 
the peak of the baryon density diminished. For an even number 
of colours, the rise of the baryon number due to a vanishing partition 
function was called {\it Fermi-Einstein  condensation} 
(FEC)~\cite{Langfeldect2010,Langfeld:2009cd}. 
For small enough spatial size, we observe that 
FEC can also occur for an odd number of colours and thus in QCD-like 
theories. 

\begin{figure}
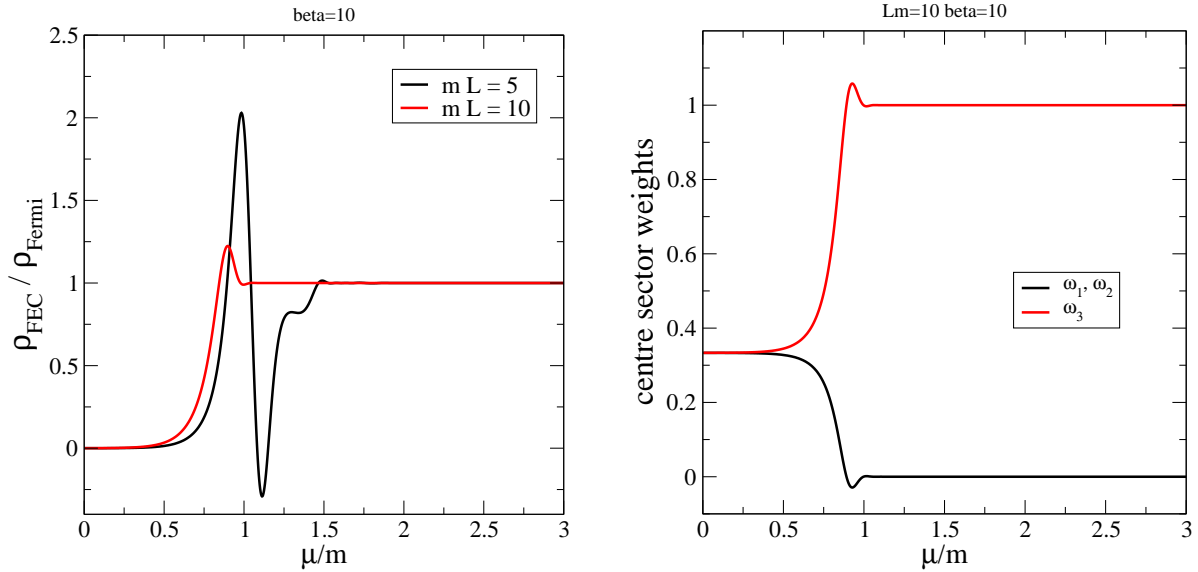

  \includegraphics[width=7.5cm]{rhoB_mu2.eps} \hspace{0.5cm}
  \includegraphics[width=7.5cm]{c_weights.eps}
\caption{\label{fig:cd2} Ratio between the baryon density of the FEC model 
and that of the standard Fermi gas model for two spatial sizes $L$, 
left panel. The real part of the centre sector weights $\omega _{1 \ldots 3}$, 
right panel. 
}
\end{figure}
\vskip0.3cm
Let us now study the baryon density for the bigger systems with 
$mL=5 $ and $mL=10$. Figure~\ref{fig:cd2}, left panel, shows the baryon
density  of the FEC model normalised to that of the Fermi gas model. 
We observe that for small values of $\mu $ the baryon density 
is suppressed compared to that from the Fermi gas model. For 
$mL=10$, this suppression ceases for $\mu/m \approx 0.9$ and 
the FEC baryon density equals that of the Fermi gas model. 
We see here confinement at work: while in the Fermi gas model, 
the baryon density rises with increasing $\mu $ by exciting 
single quarks into the system. In the FEC model at small values 
of $\mu $, the only way to increase the baryon density is to excite 
a baryon with mass $3m $ (in our model). For $\mu \approx 0.9 m$, 
deconfinement sets in and density rises further on by adding quarks 
to the system. This interpretation is corroborated by an inspection 
of the centre sector weights $\omega _i$ in figure~\ref{fig:cd2}, right 
panel. Note that $\omega _3$ corresponds to the trivial centre sector 
($z_3=1$). For small values of $\mu $, the real part of the weights 
are roughly the same and equal $\approx 1/3$. This indicates that 
centre sector transitions frequently occur wiping coloured states 
from the partition function thus installing confinement. 
For $\mu > 0.9 m $, the weights of the non-trivial centre sectors, 
i.e., $\omega _{1,2}$, vanish and only the trivial centre sector 
significantly contributes. Hence, the FEC model migrates into the 
Fermi gas model.

\subsection{Deconfinement at finite temperatures}

\begin{figure}
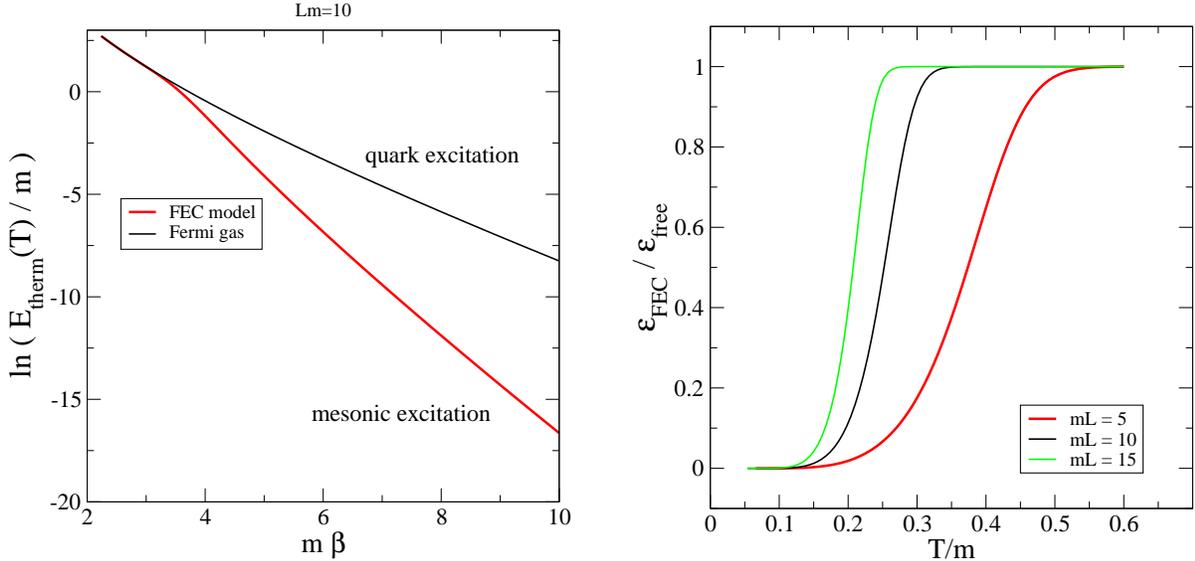

  \includegraphics[width=7.5cm]{thermal1.eps} \hspace{0.5cm}
  \includegraphics[width=7.5cm]{thermal2.eps}
\caption{\label{fig:cd3} Thermal energy $E(T)$ as a function of the 
inverse temperature $\beta $ for the FEC model and the free Fermi gas 
(left). Ratio between the thermal energy densities of the FEC model and 
the free Fermi gas (right). 
}
\end{figure}
Let us now focus on the case of vanishing chemical potential, i.e., 
$\mu = 0$. Our aim here will be to explore the phenomenology 
of the centre sector transitions far from the dense regime. 
The quantity of main interest is the thermal energy density 
$E_\mathrm{therm} (T)$ in (\ref{eq:eff22}). In the FEC quark model, 
we obtain: 
\bea 
E_\mathrm{therm} (T) &=& \sum _n \, \omega _n \; \sum _p \; 
E(p) \; \left[ \frac{ z_n^\ast }{ \E^{ \beta E(p) } + z_n^\ast }  \; + \; 
\frac{ z_n }{ \E^{ \beta E(p) } + z_n } \; \right] \; .  
\label{eq:eff35}  
\ena  
The centre sector partition functions $\Gamma  ^{(n)}_\mathrm{den}(\beta, L,
\mu=0)$ are real and positive functions of the temperature 
$T=1/\beta$. The centre sector weights $\omega _i$, $i=1 \ldots 3$ 
satisfy 
\be 
\omega _i \in \R , \hbo \omega _i \ge 0 , \hbo \sum _{i=1}^3 \omega _i 
\; = \; 1 \; , 
\label{eq:eff36}  
\en
and, thus, can be interpreted as the probability with which each sector 
contributes to e.g.~the thermal energy in (\ref{eq:eff35}). For later 
reference, we also quote the thermal energy of a free Fermi gas: 
\bea 
E_\mathrm{therm}^\mathrm{free} (T) &=&  2 \sum _p \; 
E(p) \; \frac{ 1 }{ \E^{ \beta E(p) } + 1 }  \; . 
\label{eq:eff35b}  
\ena  
Note that antiquarks contribute in the same way as quarks. Hence, 
the factor $2$ in (\ref{eq:eff35b}). 

\vskip 0.3cm 
\begin{figure}
\begin{center}
  \includegraphics[width=9cm]{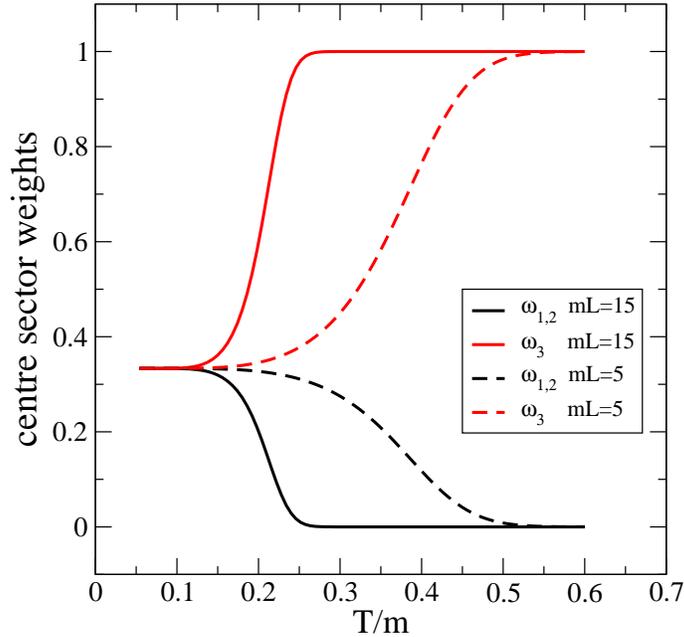} 
\end{center}
\caption{\label{fig:cd4} Probabilistic weights for the realisation 
of the centre sector $n$. 
}
\end{figure}
For first insights, we consider the case of small temperatures: 
$$ 
\beta \; m \; \ll \; 1 \; . 
$$
The partition function (\ref{eq:eff5}) can then be approximated by
\bea 
Z_Q &=& \E^{\Gamma _0} \; \sum _{n=1}^{N_c} \;  \prod _p \, 
\Bigl[ 1  +  z_n^\ast \, \E^{ - \beta E(p)} \Bigr]^2 \;  
\Bigl[ 1  +  z_n \, \E^{ - \beta E(p)} \Bigr] ^2 
\label{eq:eff37} \\
& \approx & \,3\,\E^{\Gamma _0} \, \Bigl[ 1  + \, 2 \, \sum _p 
\E^{ - \beta \, 2 \, E(p)} \Bigr] \; . 
\label{eq:eff38}
\ena 
For $\mu = 0$, the most important contribution to the low temperature 
partition function arises from mesonic states. 
Figure~\ref{fig:cd3}, left panel, shows the thermal energy 
for the FEC model in comparison to the free Fermi gas. 
At large $\beta $ (low temperatures), the FEC thermal energy 
is suppressed like $\exp\{ - \beta \, 2m \}$ since only
  mesonic excitations are possible by virtue  of confinement. This is in
  contrast to the free Fermi gas where the contribution from single quark
  states yields a suppression only of order  $\exp\{ - \beta \, m \}$. 

\vskip 0.3cm 
Dividing the thermal energy by the spatial volume yields the 
thermal energy density: 
$$ 
\epsilon _\mathrm{FEC} \; = \; E_\mathrm{therm}/L^3, \hbo 
\epsilon _\mathrm{free} \; = \; E_\mathrm{therm}^\mathrm{free}/L^3 . 
$$
Figure~\ref{fig:cd3}, right panel, shows the ratio of thermal energy densities 
of the FEC model and the free Fermi gas. 
Around $T\approx 0.2$, we observe that the FEC energy density rapidly 
approaches that of the Fermi gas. We attribute this behaviour to 
a deconfinement phase transition. Indeed, figure~\ref{fig:cd4} shows 
the centre sector weights $\omega _i$. For small temperatures $T \ll m$, 
we observe that all weights roughly equal $1/3$ indicating an 
equal contribution from the centre sectors to thermodynamical quantities. 
For $T>0.2m$, we find that 
the non-trivial centre sectors cease to contribute implying that 
the FEC model turns into the free Fermi gas theory at high temperatures.

\subsection{The phase diagram from confinement}

\begin{figure}
  \includegraphics[width=7.8cm]{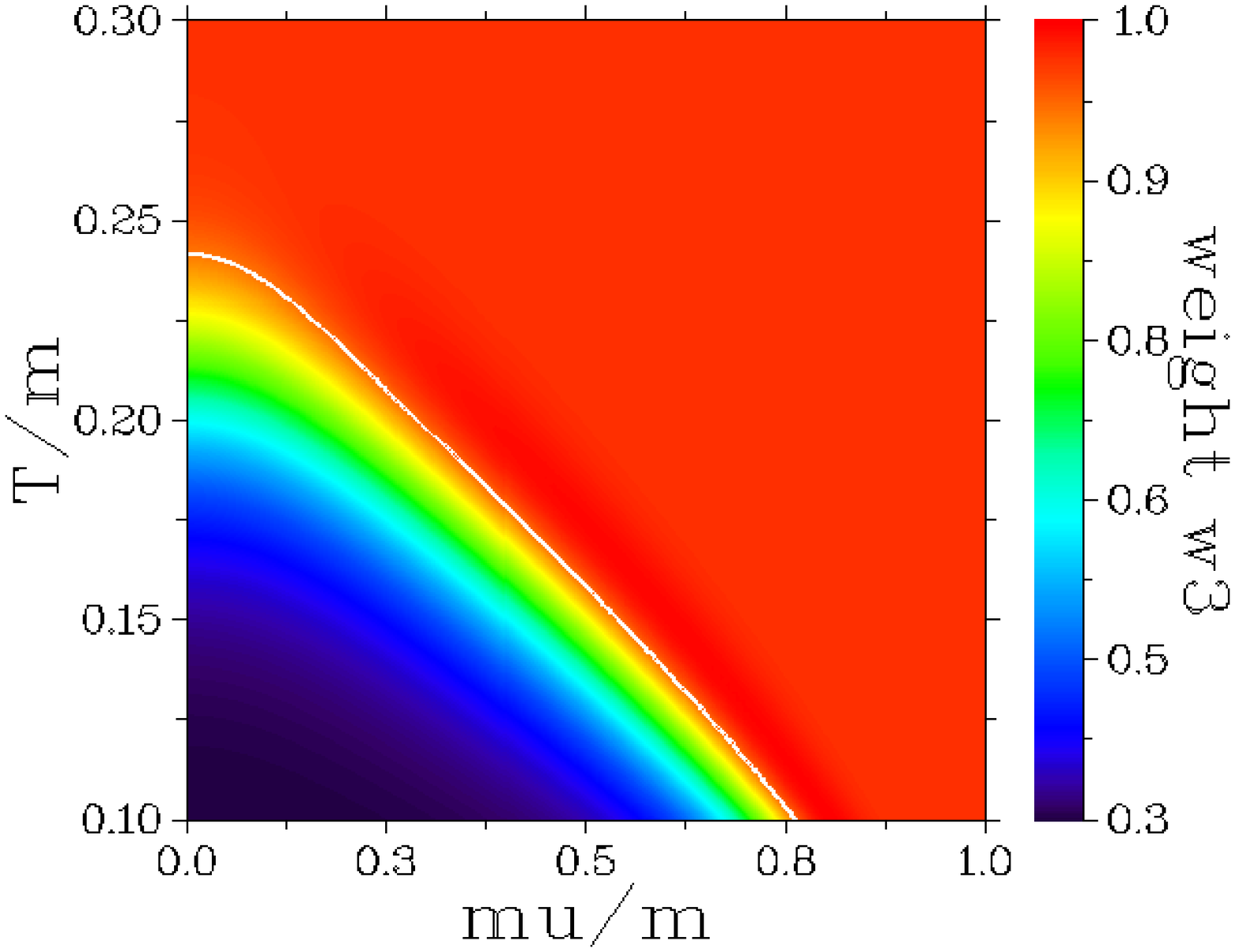} !\hspace{0.5cm}
  \includegraphics[width=7.8cm]{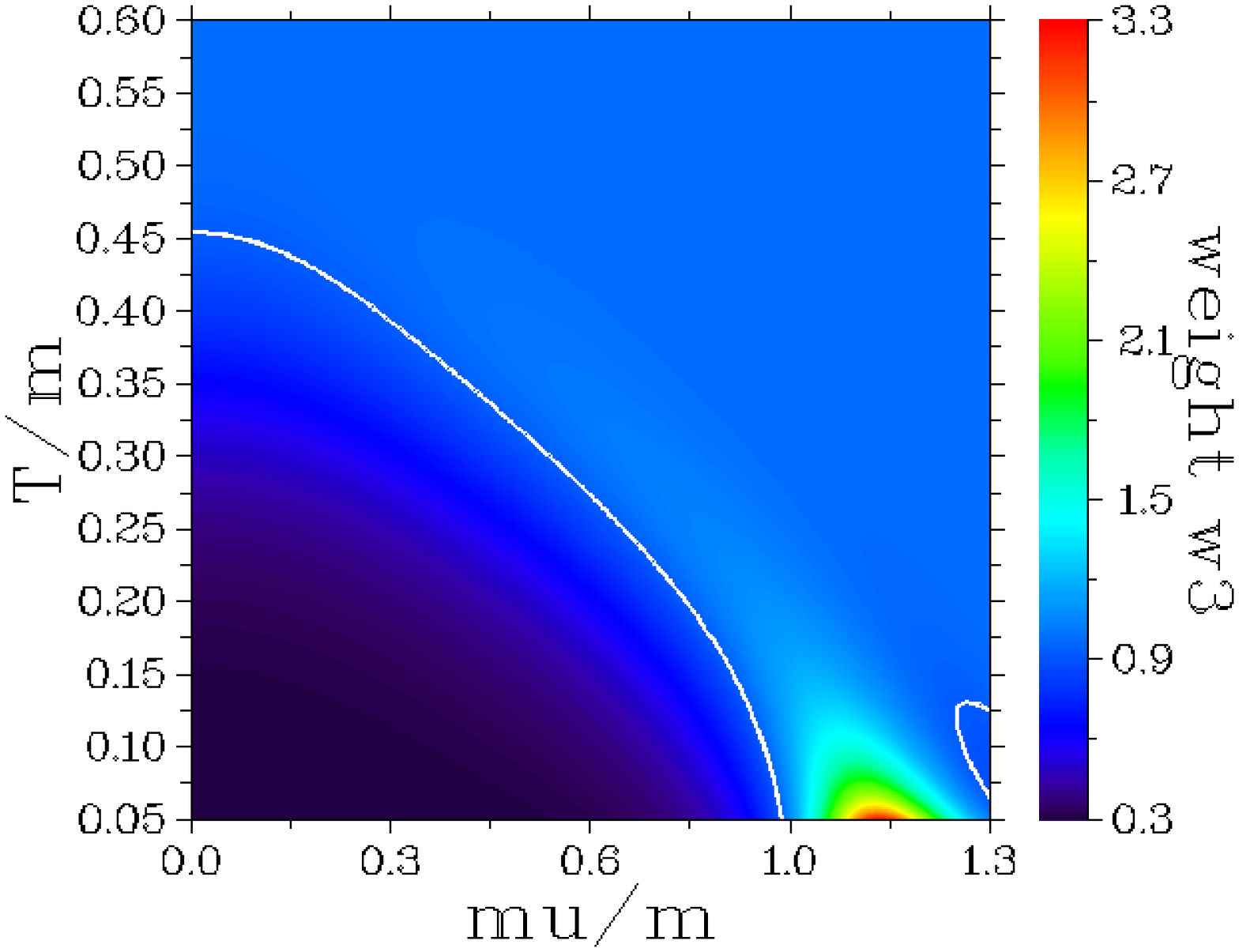}
\caption{\label{fig:cd5} Phase diagram of the FEC model as shown 
by the centre weight $\omega_3$. Left: large volumes, i.e., $mL=15$. 
Right: small volumes, i.e., $mL=5$. The contours correspond to 
$\omega _3=0.95$. 
}
\end{figure}
We are now considering both finite temperatures as well as a non-vanishing
chemical potential. In order to trace out the phase diagram, we will 
employ the centre weight $\omega _3 \in \R$: for a vanishing 
chemical potential, all weights $\omega _i$ are real numbers which sum to 
unity. At high temperatures, we have observed in the previous subsection 
that $\omega _3 \to 1$ while $\omega _{1,2} \to 0$ thus showing confinement. 
For large non-vanishing chemical potentials, the latter is still true: 
$\omega_3$ close to $1$ signals the transition of the FEC model to 
the free Fermi gas. Thus, an inspection of $\omega _3 (\mu,T)$ 
maps out the deconfinement region in the phase diagram. 

\vskip 0.3cm 
Note that the $\omega_3 \le 1$ only strictly holds for $\mu =0 $ 
where the weights $\omega _i$ enjoy an interpretation as probabilities. 
Though in the deconfinement regime $\omega_3$ will approach $1$, 
there is no need for $\mu \not=0$ that $\omega_3$ is bounded from above 
by $1$. In fact, we have seen for small spatial extensions 
$ mL \stackrel{<}{_\sim } 2.5 $ that Fermi-Einstein condensation occurs 
with potentially large values of $\omega _3$ (see figure~\ref{fig:cd2}). 

\vskip 0.3cm 
Figure~\ref{fig:cd5} shows our colour coded result for 
$\omega _3(\mu,T)$. In the large volume limit, e.g.~for $mL=15$, 
the result is as expected: there is quite a sharp transition between 
the hadronic regime at low $\mu$ and $T$ to the deconfinement regime 
under extreme conditions. For a smaller volume, e.g.~$mL=5$ (right panel), 
large values of $\omega _3$ can be observed for low temperatures 
and $\mu$ close to the mass threshold. This is the regime 
for which the partition function cancels to a large extent due to 
centre sector transitions.

\section{Centre sector transitions in the gauged Ising model \label{sec:gi} }

Above using the Schwinger model as well as a SU(3) effective quark 
theory, we have stressed the phenomenological importance of transitions 
between the centre sectors of the theory under investigations. 
It remains to show that these centre transitions do occur if dynamical 
matter is present: since this matter explicitly breaks centre symmetry, 
transitions between centre sectors could be prohibited in the 
infinite volume limit. 
With ``matter'', we here address any dynamical fields (in addition to the 
Yang-Mills gauge fields) which transform under the fundamental 
representation of the gauge group. An important example is 
the theory of strong interactions, QCD, which is an $SU(3)$ Yang-Mills 
theory with e.g.\ three (light) flavours of quark matter. 

\vskip 0.3cm 
In the remainder of the paper, we discuss theories at zero chemical 
potential and search for potential transitions between the
centre sectors. Since the Schwinger model does not possess genuine 
phase transitions due to its 2-dimensional nature, we start 
the considerations with 
the $\Z_2$ gauge theory coupled to Ising matter in 3-dimensions. 
This theory is easily accessible by means of Monte-Carlo simulations, and 
offers the possibility of phase transitions. 
We will focus on two competing scenarios 
which both sketch a different picture of the realisation of the centre 
symmetry: 

\begin{itemize}
\item[(i)] The explicit breaking triggers a spontaneous breakdown of 
centre symmetry. Centre sector tunneling does not take place. 
Observables only receive contributions from the trivial centre sector. 

\item[(ii)] The explicit breaking of centre symmetry breaking does not 
bring an end to the centre sector transitions. Observables are still 
averaged over the different sectors. There is, however, a bias towards 
the trivial sector. 

\end{itemize}

Note that, in scenario (ii), the matter sector is a correction to the 
gauge sector of pure Yang-Mills theory. The Polyakov line expectation value
is only non-zero since the lattice configurations are biased towards the 
trivial centre sector. Centre symmetry does also break spontaneously 
at high temperatures though the critical temperature can be 
quantitatively different from the pure Yang-Mills case. 
Note that the idea of a spontaneously broken symmetry which is also 
explicitly broken 
is a useful concept: in QCD, chiral symmetry is explicit broken 
by the current quark masses. Additional spontaneous chiral symmetry breaking 
assigns the pions the role of ``almost'' Goldstone bosons and explains 
their particular role in the hadron spectrum. 

\subsection{Phase diagram of the gauged Ising model } 

We are here going to discuss the $\Z_2$-gauge theory minimally coupled to 
Ising spins $\sigma _x \in \{\pm 1\}$ to ensure $\Z_2$ gauge invariance. 
The model has been considered as the most simple gauged Higgs 
theory, and it has been shown that its phase diagram has 
the same qualitative features than e.g.~the SU(2)-Higgs 
theory~\cite{Wegner:1984qt,Fradkin:1978dv,Creutz:1979he,Jongeward:1980wx,Brezin:1981zs}. 
As familiar from lattice Yang-Mills theories, the gauge fields are represented
by the links $Z_\mu (x)  \in \{\pm 1\}$. The partition function of the theory
is given by 
\bea 
Z &=& \int {\cal D}Z_\mu \; {\cal D} \sigma \; \exp \{ S_Z\} , 
\label{eq:gi1} \\ 
S_Z &=& \beta_\mathrm{I} \sum _{x, \mu >\nu} P_{\mu \nu}(x) \; + \; \kappa 
\sum _{x \, \mu} \sigma _x Z_\mu (x) \, \sigma _{x +\mu } \; , 
\label{eq:gi2} 
\ena
where $P_{\mu \nu}(x)$ is the usual plaquette
\be
P_{\mu \nu}(x) \,=\, Z_\mu (x) \; Z_\nu (x+\mu) \; Z_\mu (x+\nu) \; Z_\nu (x) \;.
\label{eq:gi3} 
\en
One easily verifies 
the invariance of the partition function under gauge 
transformations ($\Omega _x \in \{\pm 1\}$):
\be 
\sigma \to \sigma ^\Omega _x \; = \; \Omega (x) \, \sigma _x , \hbo 
Z_\mu (x) \to Z_\mu^\Omega (x) \; = \; \Omega (x) \, Z_\mu (x) \, \Omega 
(x+\mu) \; . 
\label{eq:gi4} 
\en 
For $\kappa =0$, the Ising matter decouples from the gauge sector, and 
we are dealing with a pure $\Z_2$ gauge theory. It is well known that this 
theory confines static centre charges, and the static potential 
between two static charges is linearly rising with their distance. 
The Polyakov lines is an order parameter for confinement. 
For non-vanishing but small values $\kappa $, the dynamical Ising 
spins can screen the static centre charges, and we obtain 
{\it string breaking}: the linearly rising static potential 
at large distances flattens when the potential energy is sufficient 
to create a dynamical matter pair. Hence, the phase diagram as a function 
of $\kappa $ and $\beta_\mathrm{I} $ is expected to show the same features 
as that of the SU(2) Higgs theory. 

\begin{figure}[h]
  \includegraphics[width=7cm]{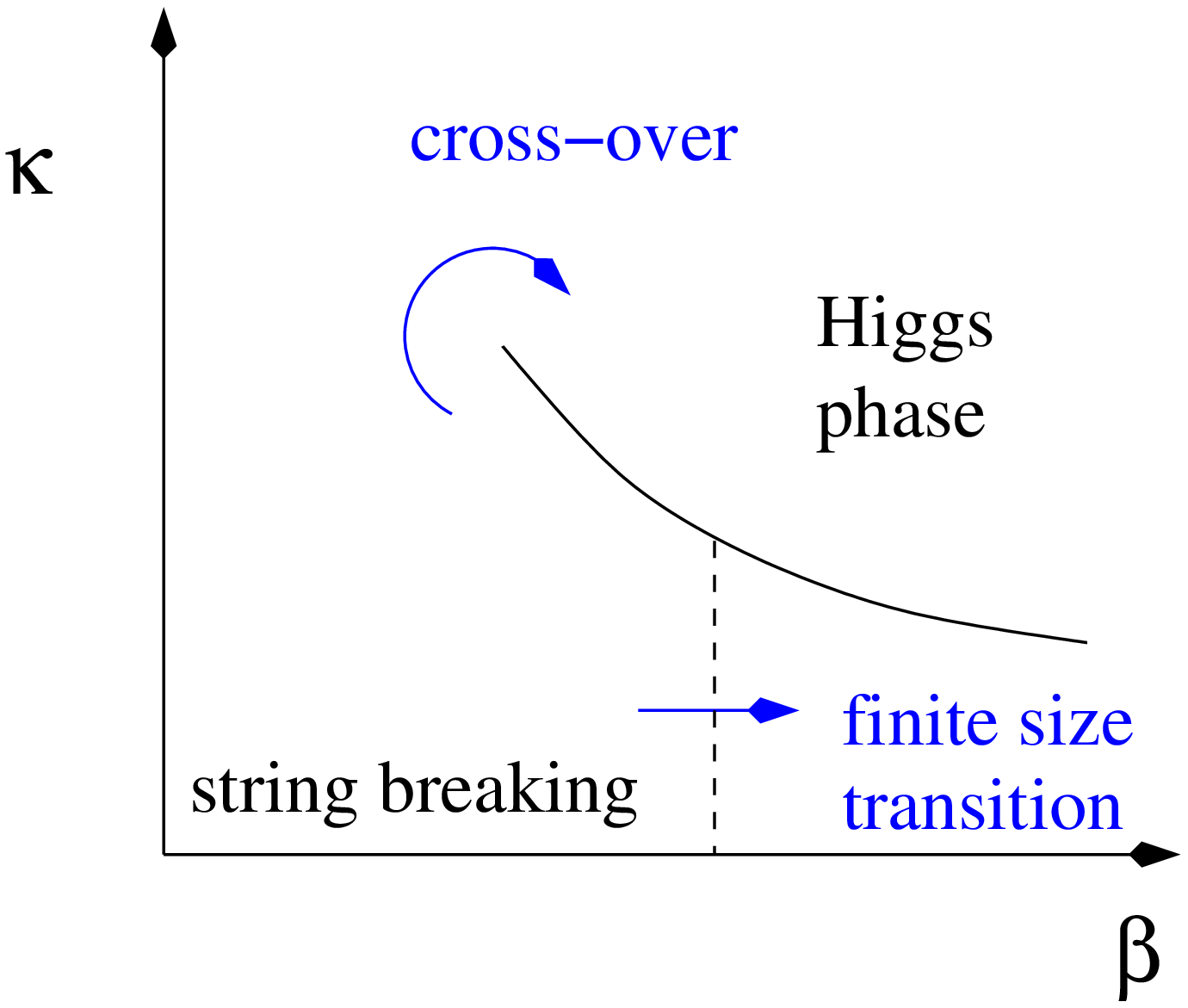}  \hspace{0.5cm}
  \includegraphics[width=9cm]{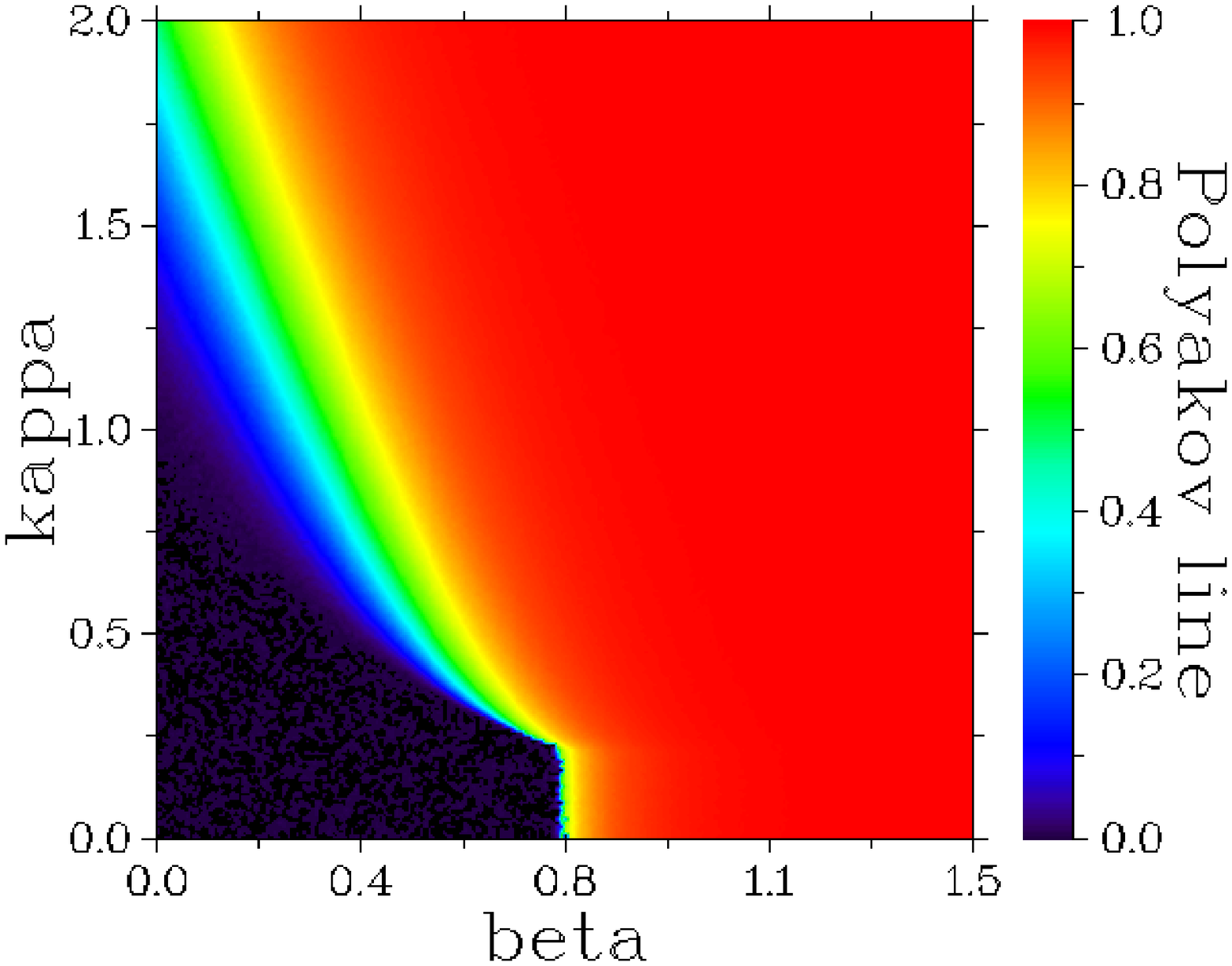}  \hspace{0.5cm}
\caption{\label{fig:gi1} Sketch of the phase diagram (left); 
Polyakov line expectation value of the gauged Ising model as 
function of $\beta_\mathrm{I} $ and $\kappa $.
}
\end{figure}
Figure~\ref{fig:gi1}, left panel, 
summarises this diagram: due to the lack of a local order parameter, 
a cross-over region from the string breaking regime to the 
so-called Higgs phase is expected. We verified these expectations by 
calculating Polyakov line expectation values using a $20^3$ lattice. 
Both, the update of the links $Z_\mu (x)$ and the matter fields 
$\sigma _x$ have been done by standard heat bath techniques. 
We calculated the Polyakov  expectation value for $204 \times 201$ 
different values in the $\beta_\mathrm{I} $-$\kappa $ plane. Our numerical 
findings are colour-coded and summarised in figure~\ref{fig:gi1}, right 
panel. Both, the cross-over regime as well as the finite size transition is 
clearly visible.

\subsection{The interface tension of the trivial centre sector }

Let us briefly discuss the ``empty'' vacuum of the gauged 
Ising model. The discussion in the context of pure Yang-Mills theory 
in subsection~\ref{sec:empty} can be straightforwardly transferred 
to the case of the $\Z_2$ gauge theory: The gauge inequivalent 
empty vacua are characterised by the values of the 
homogeneous Polyakov lines in each direction. Thus each of the Polyakov 
lines $P_1$, $P_2$ and $P_3$ takes values $\pm 1$, there are $2^3=8$ 
states which we need to consider. The $\Z_2$ \emph{centre transformation} 
\be 
Z_\nu (x) \to (-1) \, Z_\nu (x) \hbo \forall \; \; \; 
x_\mu , \; \mu \not= \nu, \hbo x_\nu \; \; \hbox{fixed} 
\label{eq:gi10}
\en 
is discrete, changes the sign of the Polyakov line $P_\nu$ and 
therefore maps one empty vacuum state to another. 

Before we proceed to consider the centre sector transitions within the 
gauged Ising model, we here discuss centre interfaces in the particular 
sector with all Polyakov lines trivial. For this ``empty'' vacuum state 
a gauge  can be found where all links are one: 
\be 
(P_1,P_2,P_3) \; = \; (1,1,1) \; , \hbo Z_\mu(x) =1 \, \; \; \; 
\forall \; \; x, \; \mu \; . 
\label{eq:gi11}
\en 
The gauged Ising model then collapses to the standard Ising model 
with ferromagnetic bonds only. 

\vskip 0.3cm 
\begin{figure}[t]
\begin{center}
  \includegraphics[width=12cm]{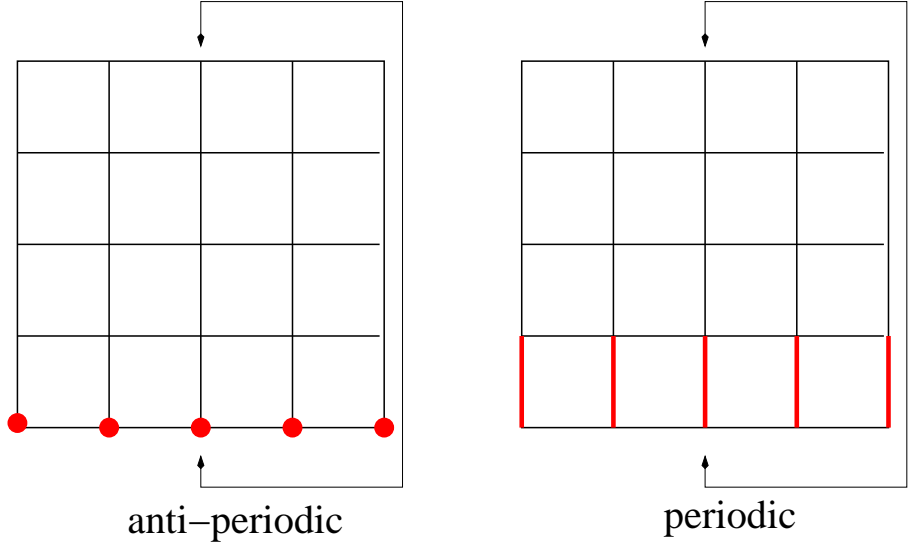} 
\end{center}
\caption{\label{fig:gi2} Partition function with anti-periodic boundary 
conditions rewritten as a partition function with a centre twist. 
}
\end{figure}
If $\vert \psi \rangle $ is particular 2d array of spin at $x_0=0$, i.e., 
the so-called {\it in}-state and if $\vert ^z\psi \rangle $ 
is a centre copy of this state for which all spin are reflected, 
we would like to investigate the overlap of the true vacuum 
state $\vert \psi _0 \rangle $ with either $\vert \psi \rangle $ {\it 
and} $\vert ^z\psi \rangle $. If centre symmetry is realised in the Wigner 
mode, the vacuum is centre symmetric yielding an overlap in both cases. 
If the centre symmetry is spontaneously broken, the overlap vanishes 
for one of the matrix elements. To investigate the spontaneous breaking 
of centre symmetry, we thus study: 
\be 
\chi \; = \; \frac{  \bigl\langle ^z\psi \big\vert 
\exp \{- H/T \}  \big\vert \psi \bigr\rangle  }{ 
\bigl\langle \psi \big\vert 
\exp \{- H/T \}  \big\vert \psi \bigr\rangle } \; , 
\label{eq:gi15}  
\en 
where $H$ is the Hamilton operator and $T$ is the temperature. 
For sufficiently small temperatures, the exponentials in the 
latter equation project onto the ground state $\vert \psi _0 \rangle$, 
and we obtain: 
\be 
\chi \; \to \; \frac{  \bigl\langle ^z\psi \big\vert \psi _0
  \bigr\rangle  \, \bigl\langle \psi _0 \big\vert \psi 
  \bigr\rangle  }{  \big\vert \, \bigl\langle \psi _0 \big\vert \psi 
  \bigr\rangle \, \big\vert ^2 }  \; = \; 
\left\{ \begin{array}{l} 
1 \; \; \; \hbox{Wigner-Weyl realisation} \\ 
0 \; \; \; \hbox{spontaneous symmetry breaking.}
\end{array} \right.  
\label{eq:gi16}  
\en 
In the functional integral approach, $\chi $ is given by the 
ratio of two partition function: the partition function with 
the spins $\sigma _x$ obeying {\it anti-periodic} boundary conditions 
constitutes the numerator in (\ref{eq:gi15}), while the standard partition function 
with period spins is in the denominator. A connection to the centre 
symmetry can be established by ``pushing out'' the centre element to 
the links: 
$$ 
^z \sigma_x \;  U_\mu (x) \; \sigma _{x+\mu } \; = \; 
\sigma_x \;  (-1) \, U_\mu (x) \; \sigma _{x+\mu } \; . 
$$
The net effect is illustrated in figure~\ref{fig:gi2}: 
\be 
\chi \; = \; \frac{ Z_\mathrm{anti-periodic} }{ Z_\mathrm{periodic}} 
\; = \; \frac{ Z_\mathrm{twist} }{ Z_\mathrm{periodic}} \; . 
\label{eq:gi17}  
\en 
We expect that the transition between centre sectors are 
exponentially suppressed with a slope given by the size of the 
centre interface. For a $N^2 \times N_t$ lattice, the minimal surface of the 
centre interface is $N^2$. It is therefore convenient to introduce 
the interface tension $\sigma $ by 
\be 
\chi \; = \; \exp \Bigl\{ - \, N^2 \, \sigma (\kappa ) \, \Bigr\} \; . 
\label{eq:gi18}  
\en 
For small values $\kappa $, $\chi $ can be directly calculated by the 
so-called strong coupling expansion, i.e., the Taylor expansion 
with respect to $\kappa $. For large values of $\kappa $, $\chi $ 
can be obtained by means of a duality transformation. In 3d Ising model 
is dual to the 3d $\Z_2$ gauge theory, and $\chi $ appears to be 
the $\Z_2$ expectation value of the maximal spatial Wilson loop 
(details will be presented elsewhere). Altogether, we find: 
\be 
\sigma (\kappa ) \; \approx \; 
\left\{ \begin{array}{ll} 
\kappa ^{N_t}  & \; \; \; \hbox{for} \; \; \; \kappa \ll 1 \; , \\ 
2 \kappa  & \; \; \; \hbox{for} \; \; \; \kappa \gg 1 \; . 
\end{array} \right.  
\label{eq:gi19}  
\en 
For a fixed aspect ratio $N_t/N$ and in the thermodynamic limit 
$N\to \infty$, we find that the theory is in the disordered phase 
for small $\kappa $, i.e., $\chi \to 1$, implying that centre sector
transitions do occur frequently. For large $\kappa $, the 
interface tension is independent of $N_t$ and large showing 
that spontaneous centre symmetry breaking occurs in this case. 

\vskip 0.3cm 
\begin{figure}[t]
\begin{center}
  \includegraphics[width=12cm]{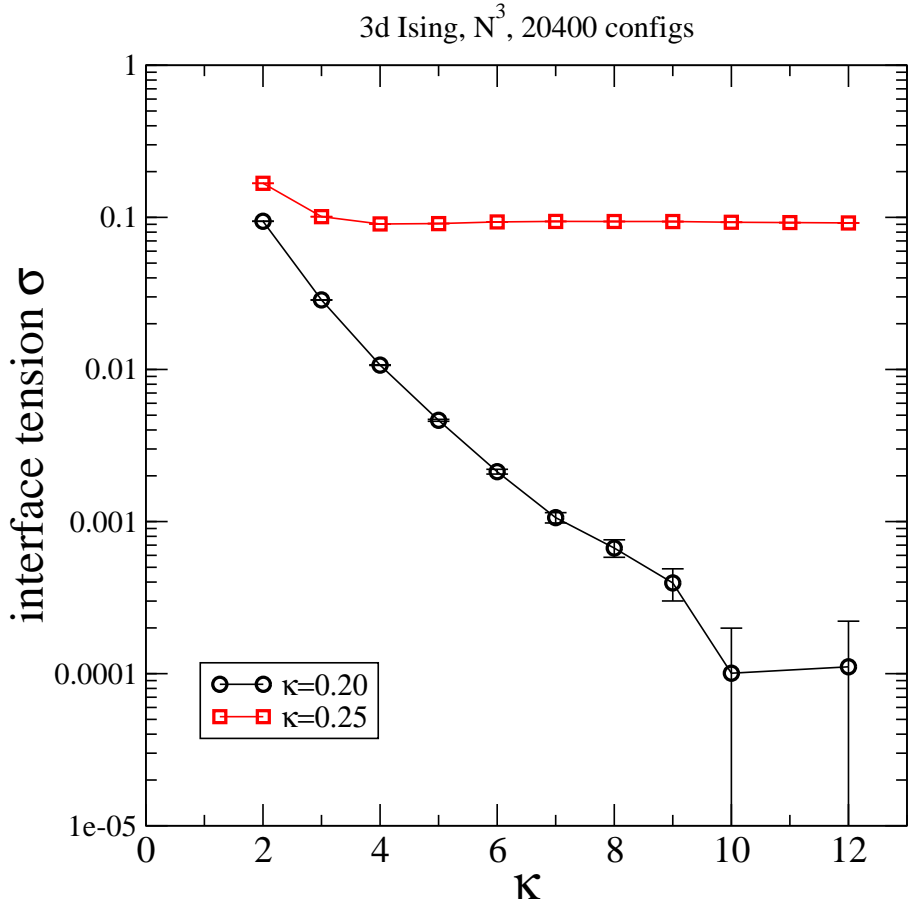} 
\end{center}
\caption{\label{fig:gi2b} The interface tension of the 3d Ising model 
as a function of the system size $N$ for interaction strength 
$\kappa =0.20 $ (disordered phase) and 
$\kappa = 0.25$ (ordered phase). 
}
\end{figure}
For intermediate values of $\kappa $ the interface tension
must be obtained by numerical means. The ratio of partition functions, 
i.e., $\chi $, can be numerically calculated in an efficient way 
using the so-called {\it snake-algorithm}~\cite{deForcrand:2004jt}. 
Our numerical results are shown in figure~\ref{fig:gi2b}. 
In the ordered phase at $\kappa =0.20$, we observe an exponential 
decrease of the interface tension $\sigma $ with $N$ in line 
with result from the leading order Taylor expansion in $\kappa $. 
For the ordered phase at $\kappa =0.25$, we confirm that 
the interface tension is indeed independent of the volume.

\subsection{Centre sectors and parametric transition }

While in the previous subsection, the link variables have been 
frozen to the trivial centre sector, we now consider the 
transition element $\chi $ of the gauged Ising model where all 
links $Z_\mu (x)$ are dynamical. Defining the twisted link variables by 
\bea 
Z_3 ^\mathrm{twist} (x) &=& (-1) Z_3(x) \hbo \forall x_1, \, x_2 \, ; \; \; \; 
x_3 \; \; \hbox{fixed} \; , 
\label{eq:gi20} \\ 
Z_\mu ^\mathrm{twist} (x) &=&  Z_\mu (x) \hskip19mm \hbox{else}, 
\nonumber 
\ena 
the partition function of the twisted partition function is given by 
\bea 
Z_\mathrm{twist} &=& \int {\cal D}Z_\mu \; {\cal D} \sigma \; \exp \{ S_Z\} , 
\label{eq:gi21} \\ 
S_Z &=& \beta_\mathrm{I} \sum _{x, \mu >\nu} P_{\mu \nu}[Z] \; + \; \kappa 
\sum _{x , \mu} \sigma _x Z_\mu ^\mathrm{twist} (x) \, \sigma _{x +\mu } \; . 
\label{eq:gi22} 
\ena 
Given the invariance of the measure and of the plaquette, 
$$ 
{\cal D } Z_\mu  \; = \; {\cal D} Z^\mathrm{twist}_\mu \; , \hbo 
P_{\mu \nu}[Z] \; = \; P_{\mu \nu}[Z^\mathrm{twist}] \; , 
$$
one easily obtains
\be 
Z_\mathrm{twist} \; = \; Z_\mathrm{periodic} \hbo \Rightarrow \hbo 
\chi \; = \; 1 \; . 
\label{eq:gi23} 
\en
In the gauged Ising model (as well as in the Schwinger model and QCD), 
the centre twist is part of the gauge field average. It is therefore,  
strictly speaking, impossible to consider a spontaneous breakdown of 
the centre symmetry. Note, however, that the matter fields break 
the centre symmetry {\it explicitly}. Depending on the coupling strength 
$\kappa $ and the temperature, $N_t$ respectively, this explicit breaking 
can be small. Actually for small $\kappa $, we are going to show that at low 
temperatures the explicit breaking is rather irrelevant while 
above a critical value for the temperature, the explicit breaking is strong 
in the sense that it does not makes sense to even consider an 
approximate centre symmetry. We call this a {\it parametric transition}.

\vskip 0.3cm 
Let us firstly illustrate the explicit breaking in the gauged Ising model. 
To this aim, we integrate over the Ising spin $\sigma _x$ for a given 
link distribution $Z_\mu (x)$. The resulting probabilistic factor 
contains contractible Wilson loops $W[Z]$ as well as the Polyakov line 
$P[Z]$ (for simplicity we here consider infinite space)
\be 
\int {\cal D} \sigma \; \exp \{ S_Z [Z] \} \; = \; 
\sum _{\cal C} \kappa ^{L({\cal C})} \; W_{\cal C} [Z] \; + \; 
\kappa ^{N_t} \sum _{x_1, x_2} P[Z] \; + \; {\cal O}\left(\kappa ^{N_t+3} 
\right) \; . 
\label{eq:gi24} 
\en 
The sum over the closed loops ${\cal C}$ is centre symmetric while the terms
containing the Polyakov loop explicitly break the symmetry. 

\vskip 0.3cm 
\begin{figure}[t]
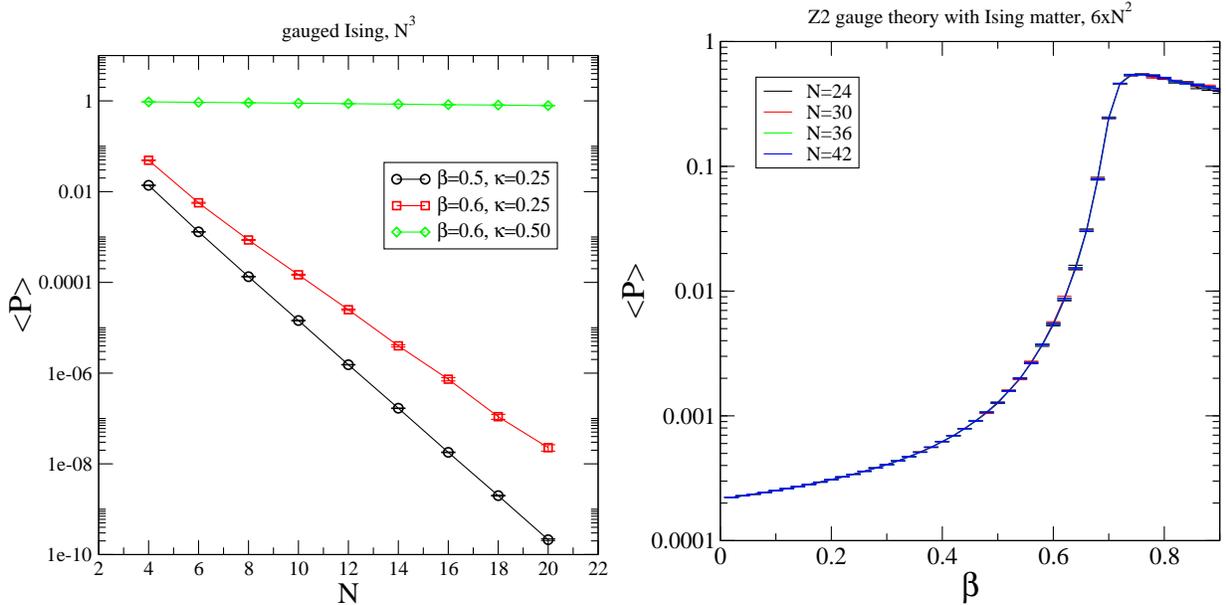

\begin{center}
  \includegraphics[width=8cm]{pol_conf.eps} 
  \includegraphics[width=8cm]{pol_z2_6N.eps} 
\end{center}
\caption{\label{fig:gi3} Polyakov line expectation values of the gauged 
$N^3$ Ising model as a function of the system size $N$ (left). 
Same quantity for a $6\times N^2$ lattice as a function of $\beta_\mathrm{I} $
(right).  
}
\end{figure}
In following, we study the theory by means of Monte-Carlo methods. 
We used a heat-bath update for the link and matter fields. 
Since the matter action describing 
the gauge field matter interactions is local, we used a 
generalised Luescher-Weisz algorithm~\cite{Luscher:2001up,Luscher:2002qv}
for the calculation of the Polyakov line expectation value. 

\vskip 0.3cm 
We are now considering moderate values for $\kappa $, let us say 
$\kappa \le 0.25$. 
We have already obtained (see figure~\ref{fig:gi1}) 
that for small $\beta_\mathrm{I} $ the Polyakov line is small. Increasing
$\beta_\mathrm{I} $  above a critical value, the Polyakov line expectation
value  
$\langle P \rangle $ rapidly takes large values. 
In fact, we find that, below the critical value, 
$\langle P \rangle $ is strictly vanishing in the infinite 
volume zero temperature limit (for fixed and finite aspect ratio) 
although $\langle P \rangle \not=0$ for any finite system. 
This is illustrated in figure~\ref{fig:gi3}, left panel: the graph 
shows $\langle P \rangle$ for a $N^3$ lattice as a function of $N$. 
At least in the string-breaking phase for sufficiently small $\kappa $, 
we observe an exponential decrease of $\langle P \rangle$ with 
increasing $N$. In the Higgs phase (here for $\beta_\mathrm{I} = 0.6$ and 
$\kappa = 0.30$), we still observe a slight decrease of $\langle P \rangle$  
with the system size. Whether the Polyakov line expectation finally 
vanishes for large $N$ cannot be inferred from the present data. 
Further investigations are left to future studies. 

\vskip 0.3cm 
To monitor the deconfinement cross-over at 
finite temperatures, we have calculated $\langle P \rangle$ 
for a $6\times N^2$ lattice as a function of $\beta_\mathrm{I} $ for the 
aspect ratios $4,5,6,7$. Our numerical result is shown in 
figure~\ref{fig:gi3}, right panel. We observe a maximum of 
$\langle P \rangle$ for $\beta _\mathrm{I}= \beta _c  \approx 0.75(5)$ and
very little  
dependence on the system size $N$. We argue that $\beta_\mathrm{I} < \beta _c$, 
the explicit breaking of centre symmetry is weak, the physics of the 
regime might be similar to that of the symmetric theory for $\kappa =0$, and 
that perturbation theory with respect to $\kappa $ yields reliable 
results. On the hand for $\beta_\mathrm{I} \ge \beta_c$, centre symmetry
breaking  
is strong and any similarities with the symmetric theory are lost although 
we always have $\chi =1 $ in (\ref{eq:gi23}). 
This is the parametric transition.

\section{Centre-sector transitions in the $SU(2)$ Higgs
  theory \label{sec:higgs} }

\subsection{String breaking }

In order to study centre sector transitions in a more realistic, i.e., 
more QCD-like, theory, we will study a theory where the''empty vacuum'' 
features flat directions, where this vacuum symmetry collapses to 
the discrete centre symmetry upon the inclusion of quantum fluctuation 
and where the trivial centre-sector is favoured through the matter 
sector. In addition and for the first time in this paper, we will consider 4
Euclidean  space-time dimensions. 

\vskip 0.3cm 
The most simple such theory is a $SU(2)$ gauge theory with a scalar 
field (say Higgs) in the fundamental representation. This theory is 
accessible by means of lattice gauge theories with large statistical
accuracy. The partition function is given by 
\bea 
Z  &=& \int {\cal D}U \; {\cal D}\phi \; {\cal D}\phi ^\dagger \; 
\exp \{ S_\mathrm{Wil} + S_\mathrm{Higgs} \} \; , 
\label{eq:qv10}  \\ 
S_\mathrm{Higgs} &=& \kappa \, \sum _{x, \mu} \; \mathrm{Re} \, 
\phi ^\dagger (x) \, U_\mu (x) \, \phi (x+\mu) \; - \; 
\sum _x  \; \Bigl[ \frac{1}{2} \, \phi^\dagger (x) \, \phi(x) \; + \; 
\lambda \; [\phi^\dagger (x) \, \phi(x) ]^2 \Bigr] \; . 
\nonumber
\ena
The parameter $\kappa $ quantifies the interaction strengths of the scalar
fields with the gauge sector, while $\lambda $ gives rise to a quartic Higgs
self-interaction.  

\vskip 0.3cm 
\begin{figure}
  \begin{center} 
  \includegraphics[width=10cm]{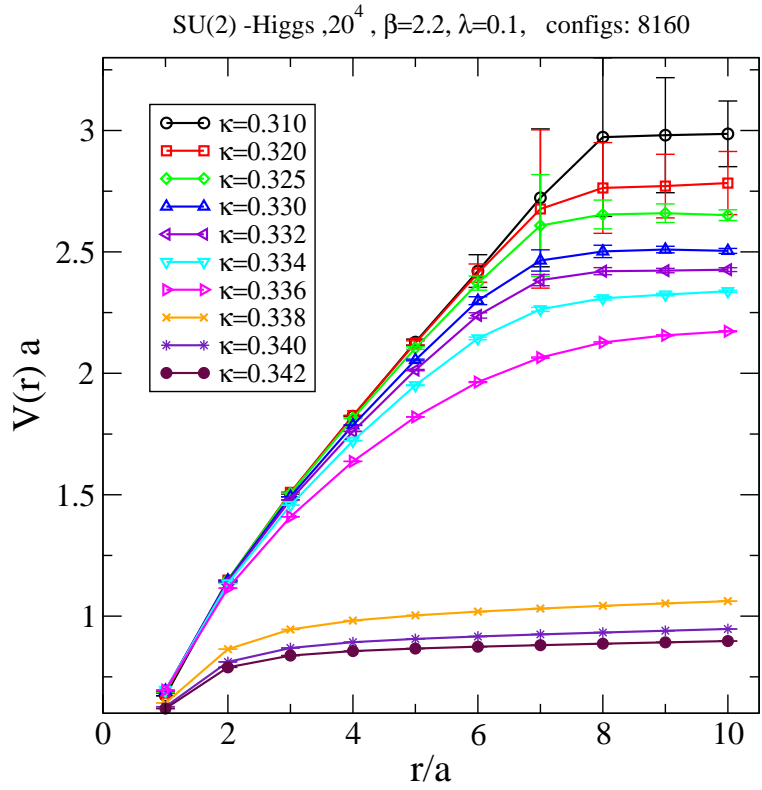} 
  \end{center} 
\caption{\label{fig:4} For the SU(2) Higgs theory ($\lambda =0.1$) 
  the static quark antiquark potential for a $20^4$ lattice,
  $\beta_\mathrm{Wil} =2.2$ and several values for $\kappa $, 
 $8160$ configurations per set.
}
\end{figure}
The SU(2)-Higgs theory shares the so-called {\it string breaking} 
with QCD: at low temperatures and sufficiently weak gauge-matter 
interactions, the static quark anti-quark potential linearly rises 
at intermediate distances between the static quark antiquark pair while it 
flattens at asymptotic distances. A popular picture assumes that 
a colour electric flux tube still forms between a static quark antiquark 
pair and that this string breaks due to the creation of a 
Higgs anti-Higgs pair. Once the string is broken, there is little penalty 
in energy to increase any further the distance between the (screened) heavy
quarks. The static potential approaches a constant. 
This also implies that the Polyakov line expectation is non-zero 
(even at low temperatures) and cannot serve any more as an order 
parameter. 

\vskip 0.3cm 
The phase diagram has been qualitatively described in figure~\ref{fig:gi1}. 
On a more quantitative note, we have calculated the static potential 
by means of lattice gauge simulations. In order to ensure 
good ergodicity, we used the {\it local hybrid Monte-Carlo} update for 
the link fields as well as the scalar fields. In this paper, we work with
$\lambda =0.1$ and $\beta_\mathrm{Wil} \in [2.2 \ldots 2.5]$ which 
roughly corresponds to the scaling window, at least, at small to moderate 
values~$\kappa $. 

\vskip 0.3cm 
To calculate the static quark potential, we here followed the two channel
approach developed  in~\cite{Philipsen:1998de,Knechtli:1998gf}. 
Using a $20^4$ lattice, our results for the Wilson coupling
$\beta_\mathrm{Wil} =2.2$,  
fixed quartic coupling  $\lambda =0.1$ and several values of $\kappa $ 
is shown in figure~\ref{fig:4}. For each potential, $8160$ independent 
lattice configurations contributed the relevant expectation values. 
For small values of $\kappa $, e.g.~$\kappa < 0.336$,  the string breaking
scale can be intuitively set by an inspection of the  graph: for $\kappa =
0.31$, we would set $r_b \approx 7 \, a$. For larger  values of $\kappa $, a
gradual transition to the Higgs phase sets in, and a string breaking scale is
hardly defined. It seems, however, that  
this scale cannot be pushed to small values by just altering the Higgs 
hopping parameter $\kappa$. If we fit the static potential for 
$r<r_b$ to $V(r) = a_0 + \sigma_\mathrm{int} r + a_1/r $, we define the so-called 
intermediate string tension $ \sigma_\mathrm{int}$. 
For $\kappa =0.31$, our estimate is 
\be 
 \sigma_\mathrm{int} a^2 \approx 0.22(1) \, , \hbo r_b \, 
\sqrt{ \sigma_\mathrm{int}} \approx 3.3 \, 
, \hbo r_b \approx 7.2 \, \mathrm{fm} \; , 
\label{eq:sb1}  
\en 
if we assume the QCD value  $\sigma_\mathrm{int}\approx (440 \,
\mathrm{MeV})^2$ to set the scale. 
We also point out that for $\kappa $ values close to the transition line, 
e.g., for $\kappa = 0.336$, the string breaking scale is difficult to define. 
Here, the picture of a well defined short string might loose its validity.

\subsection{Volume dependence of the Polyakov line expectation value}

\begin{figure}
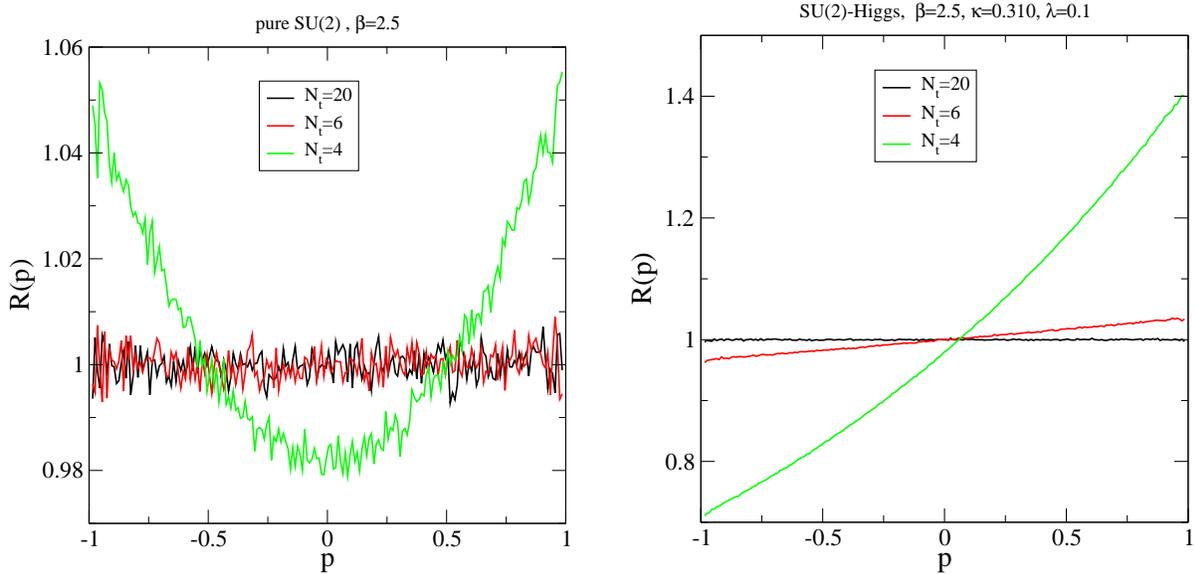

  \includegraphics[width=7.5cm]{pol_dis_YM.eps}  \hspace{0.5cm}
  \includegraphics[width=7.5cm]{pol_dis.eps} 
\caption{\label{fig:3b} Ratio $R(p)$ of the Polyakov line distribution function 
over the Haar measure distribution for three different temperatures 
in pure Yang-Mills theory (left) and for the SU(2) Higgs theory 
on the $20^3 \times N_t$ lattice (right).  
}
\end{figure}
Integrating over the Higgs fields for a fixed link background yields an 
effective action for the links which is not anymore centre-symmetric. 
In order to get a first impression of the amount of {\it explicit breaking}
we consider the normalised probability distribution for the 
spatial average of the Polyakov line,
\be 
W[p] \; = \left\langle \, \delta \left( p  -  \bar P[U] \right) \, 
\right\rangle \; , 
\label{eq:qv6} 
\en 
where the expectation value is with respect to the partition function 
(\ref{eq:qv5}) and $\bar P[U]$ is the spatial average of the trace of the 
Polyakov line: 
\be 
\bar P[U] \; = \; \frac{1}{V_3} \sum _{\mbx} \; \frac{1}{N_c} \,  \tr 
\; \cP(\mbx) \; . 
\label{eq:qv6a} 
\en 
Since in the low temperature phase the probability distribution of Polyakov 
line $\cP$ is basically given by the Haar measure of the gauge group, 
the above distribution is divided by the reduced Haar measure distribution, 
\be 
R(p) \; = \; W(p) / W_0(p) \; , \hbo 
W_0(p) \; = \; \frac{2}{\pi } \, \sqrt{1 - p^2} 
\label{eq:qv7} 
\en 
to extract any effects beyond the trivial distribution. 

\vskip 0.3cm 
Let us first study pure Yang-Mills theory without Higgs matter. 
Our result is shown in figure~\ref{fig:3b}, left panel.  
At zero temperatures, i.e., $N_s=N_t=20$, for $\beta_\mathrm{Wil} = 2.5$, 
this ratio is basically independent of $p$. We increase the 
temperature $T=1/N_t \, a$ by decreasing the extent of the lattice 
in time direction. For $N_t=6$ at $\beta_\mathrm{Wil} =2.5$, pure Yang-Mills
theory is  
still in the confinement phase. Here we observe that deviations from 
the Haar measure distribution is still marginal. 
For $N_t=4$, Yang-Mills theory is in the high temperature deconfinement 
phase, and significant deviations from the Haar measure distributions 
are clearly visible for large values for $p$. Here, a bias towards the 
centre elements, $1$ and $-1$, is seen. In the infinite volume limit, 
tunneling between the centre sectors ceases to take place 
leading to a spontaneous breakdown of centre symmetry. 
It is, however, important to notice that the $N_t=4$ distribution $W(p)$ 
is to a good extent symmetric under the reflection $p \to -p$ 
by virtue of the centre invariance of the functional integral. 
This signals good ergodicity of the algorithm even 
in the high temperature phase. 

\vskip 0.3cm 
Let us now consider the SU(2) Higgs theory. 
We here used a $20^3 \times N_t$ 
lattice and $\beta_\mathrm{Wil} = 2.5$, $\lambda =0.1$ and $\kappa =0.31$
which is somewhat below the critical value for the transition from the 
confining phase to the Higgs phase. 
Figure~\ref{fig:3b}, right panel, summarises our findings.  
At very low temperatures, i.e., for $N_t=20$, there is hardly any 
deviation from the Haar measure distribution visible. 
Still in the string-breaking phase at $N_t=6$, there is 
a bias towards positive values of the Polyakov line noticeable. 
At high temperatures, such as for $N_t=4$, above the deconfinement 
transition, this tendency is strongly amplified: the centre sector 
$z_1=-1$ is even more suppressed while the probability for positive 
values of $p$ reaches large values. In all cases 
(though less visible for $N_t=20$), 
the centre symmetry $U \to U^c$, (\ref{eq:qv2},\ref{eq:qv3}) 
is {\it explicitly} broken, and the Polyakov line probability distribution 
is no longer reflection symmetric, $W(p) \not= W(-p)$. 

\vskip 0.3cm 
\begin{figure}
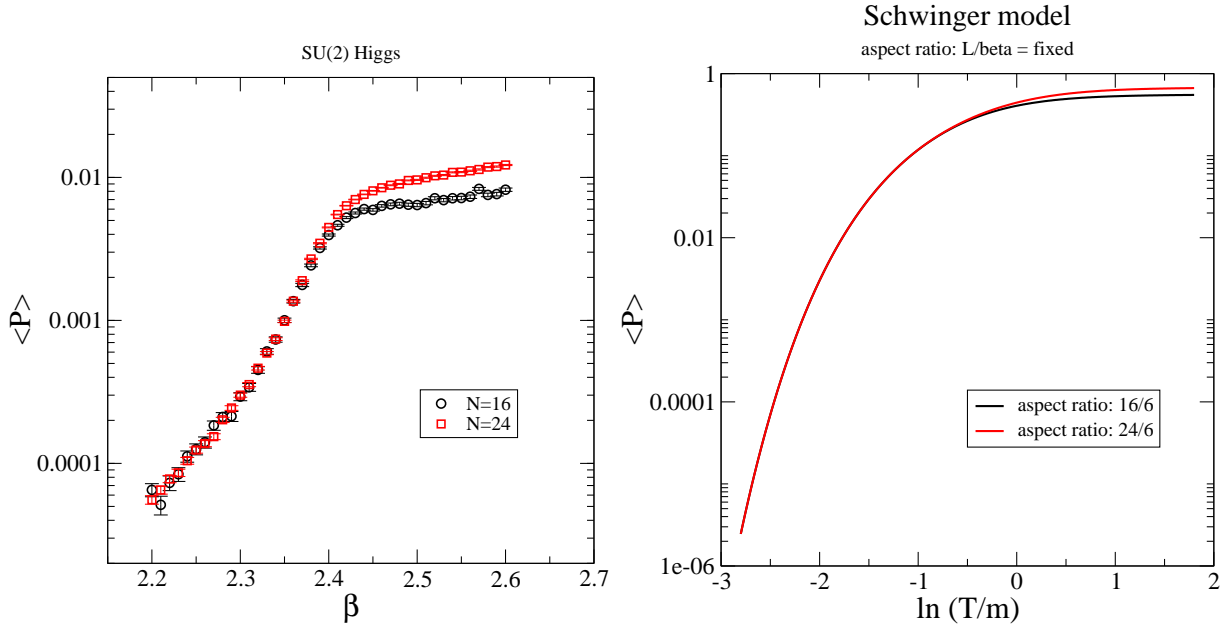

  \includegraphics[width=8cm]{pol_beta.eps} 
  \includegraphics[width=8cm]{pol_schw.eps}  
\caption{\label{fig:4c} For a fixed aspect ratio $N/N_t$, 
the dependence of the Polyakov line expectation on 
Wilson $\beta_\mathrm{Wil}$ in the SU(2)-Higgs theory 
$\kappa = 0.31$, $\lambda =0.1$, $2040$ configurations per data point, 
(left panel). Same quantity as a function of $\ln T$ in the Schwinger model 
(right panel). 
}
\end{figure}
Apparently, the explicit centre symmetry breaking is much stronger 
at high temperatures than in vacuum. Does this mean that
centre symmetry breaking is spontaneously broken 
at high temperatures? To answer this question, we have studied the 
the Polyakov line expectation value $\langle P \rangle $ for a fixed and 
finite aspect ratio, i.e., $N / N_t $, as a function of the system size. 
This can be easily achieved by using of fixed number of lattice points, 
$16^3 \times 6$ and $24^3 \times 6$, respectively, and to vary the Wilson
$\beta_\mathrm{Wil} $. The logarithm of the  
temperature $T$ in relation to the fundamental renormalisation group 
scale such as the intermediate string tension $\sigma $ 
is then roughly given by 
$$
\ln \left( \frac{T}{\sqrt{\sigma } } \right) \; = \; \ln \left(\frac{1}{N_t \, 
\sqrt{\sigma } a(\beta_\mathrm{Wil} ) } \right) \; \approx \; \gamma _1 \,
\beta_\mathrm{Wil} \;  
+ \; \hbox{constant} ,  
$$
where $\gamma _1$ is 1-loop Gell-Mann Low coefficient. For a pure 
SU(2) gauge theory, this coefficient would be $\gamma _1 = 3 \pi^2 /11 $. 
Figure~\ref{fig:4c}, left panel, shows $\langle P \rangle $ as a function of
Wilson  $\beta_\mathrm{Wil} $ using the  L\"uscher-Weisz method 
~\cite{Luscher:2001up,Luscher:2002qv} which is easily generalised to 
include the Higgs field. For small values  
of $\beta _\mathrm{Wil}$, i.e., $\beta _\mathrm{Wil}<2.4$, we roughly 
observe an exponential decrease of $\langle P \rangle $ with decreasing 
$\beta _\mathrm{Wil}$. At high values, i.e., $\beta _\mathrm{Wil}
\stackrel{>}{_\sim } 2.4$,  
$\langle P \rangle $ increases with increasing $\beta_\mathrm{Wil} $ at a
modest pace.  
Whether in this regime the dependence of $\langle P \rangle $ is 
still exponential (just with a much smaller slope) or whether 
the characteristic of this dependence has fundamentally changed cannot be 
concluded given the present set of data. 

\vskip 0.3cm 
In pure SU(2) Yang-Mills theory and a $16^3\times 4$, we would expect 
the deconfinement phase for $\beta_\mathrm{Wil}  \stackrel{>}{_\sim } 2.4$. 
It is tempting to conclude that the change of the dependence of 
$\langle P \rangle $ on $\beta_\mathrm{Wil}$ in this regime signals a
spontaneous  
breaking of centre symmetry. We here argue that this conclusion 
is premature: figure~\ref{fig:4c}, right panel, shows $\langle P \rangle $ 
as a function of $\ln T/m_\gamma  $ (where $m_\gamma $ is the induced 
photon mass which sets the scale) for the same fixed aspect ratio. 
Both curves are strikingly similar. While there is certainly 
no spontaneous breakdown of centre symmetry in the Schwinger model 
(this would lead to a Silver-Blaze problem as we pointed out in
section~\ref{sec:Schw}), the large values of $\langle P \rangle $ 
merely indicates an overlap problem if we wish to address this model 
by means of Monte-Carlo methods.

\subsection{Order parameter for centre-sector transitions }

If we consider for the moment a given lattice configuration $\{U\}$
in pure Yang-Mills and its centre copy $\{^z U\}$, then these 
configurations are degenerate in action. 
If $\{U\}$ is an ``empty vacuum'' state, i.e., 
all contractible Wilson loops on the lattice yield the unit element, 
a path in configuration space can be found which deforms 
$\{U\}$ into $\{^z U\}$ without changing the action. 
For a generic lattice configurations, this no longer the case, and 
the crucial question is whether transitions between centre-sectors 
occurs at all in the infinite volume limit. The situation is aggravated 
if dynamical matter (transforming under the fundamental representation 
of the gauge group) is present which bias the trivial centre sector. 
We have partially answered this question by studying the volume 
dependence of the Polyakov line: in the string-breaking phase, the 
bias becomes negligible in the infinite volume limit, centre transitions 
do occur and the Polyakov line averages to zero;  in the 
Higgs-phase explicit centre-symmetry breaking is strong and 
independent of the volume. 

\vskip 0.3cm 
\begin{figure}
  \begin{center} 
  \includegraphics[width=8.4cm]{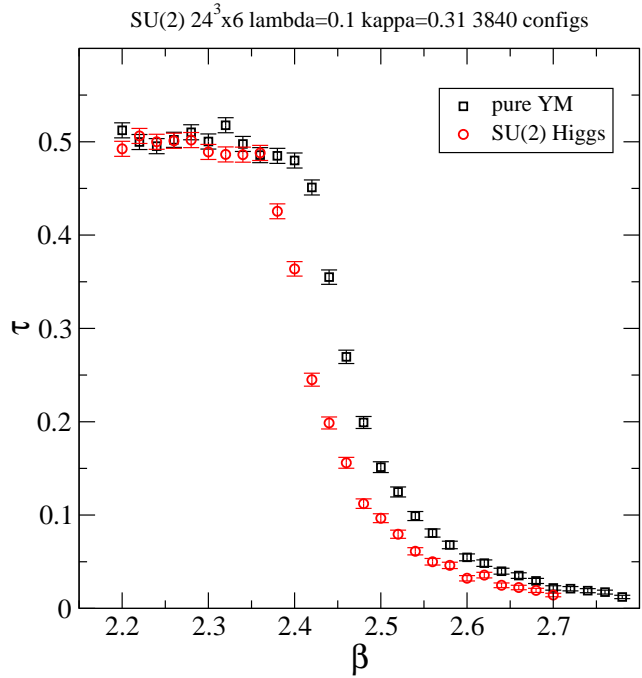} 
  \end{center} 
\caption{\label{fig:7} The tunneling coefficient for pure Yang-Mills 
theory and the SU(2) Higgs theory.}
\end{figure}
Our aim here will be to construct a sort of order parameter which 
is sensitive to the centre transitions. We do not expect that 
such an observable is built from local field operators and that it 
is an order parameter in the strict thermodynamics sense. 
It must necessarily be a non-local object which, however, nevertheless 
signals whether swapping the centre sectors can occur. 
To this aim, we divide the 3-volume into two parts of equal size, 
$V_L$ and $V_R$ 
and define the spatial average of the Polyakov line over each of the 
volumes: 
\be 
\bar P_{L/R} \; = \; \frac{1}{V_{L/R}} \, \sum _{\mbx\in V_{L/R}} \, P(\mbx) . 
\label{eq:sb4}
\en 
It is straightforward to assign a centre sector to each of 
the sublattices by the mapping 
\be 
C(\bar P) = n , \hbo n: \; \Big\vert \; \mathrm{arg} (\bar P)
- \frac{2 \pi n }{N_c}  \, \Big\vert \to \mathrm{min} \; , 
\label{eq:sb5}
\en 
where 
$$ 
\mathrm{arg} (\bar P) = \varphi \in \,]0, 2\pi] \; , \hbo 
\bar P\; = \; \vert \bar P \vert \, \exp \{ \I \varphi \} \; . 
$$
Let us then constrain the configurations in such a way that 
the left hand part of the universe belongs to centre-sector $n$ while 
the right hand part has a reference to centre sector $m$. 
The ratio of partition functions between the constrained and the 
un-constrained theory defines the free energy $F_{nm}$ for the mixing of 
the centre-sectors: 
\bea 
p_{nm}:= \exp \{ - F_{nm}/T \} &=& \frac{1}{N} \int {\cal D} U \; {\cal D}
\phi \;  {\cal D} \phi ^\dagger \; \delta \Bigl( n, C(\bar P_L) \Bigr) \; 
\; \delta \Bigl( m, C(\bar P_R) \Bigr) \; \E^{S} \; ,  
\label{eq:sb6} \\ 
N &=& \int {\cal D} U \; {\cal D} \phi \; 
{\cal D} \phi ^\dagger \;  \E^{S} \; . 
\nonumber 
\ena 
The matrix $p_{nm}$ can be interpreted as the probability to find sector $n$
in the left half and sector $m$ in the right half of the spatial universe. 
If the centre-symmetry is weakly broken in the string-breaking phase 
(or not broken at all in the infinite volume limit), the free energy 
$F_{nm}$, $n\not =m$ is finite (or might even tend to zero for an 
increasing system size), while the free energy diverges if the centre symmetry 
is strongly broken broken as e.g.~in the Higgs phase or at high
temperatures. This is basically due to the fact that the whole  
universe belongs to one centre sector, and configurations 
with $n \not=m $ have a very low probability. 

\vskip 0.3cm 
Still measuring the free energy $F_{nm}$ is hardly an easy task. 
Alternatively, we consider another intuitive measure for 
centre sector transitions: the probability $\tau $ that 
the volumes $V_L $ and $V_R $ belong to {\it different} sectors. 
We call $\tau $ the transition coefficient. 
It is directly related to the free energy by 
\bea 
\tau &=& \; \sum _{n\not=m} p_{nm} \; = \; 1 \; - \; \sum_n p_{nn} \; = \; 
1 \; - \; \sum _{n} \exp \{ - F_{nn}/ T \}  \; . 
\label{eq:sb7} 
\ena 
In the infinite volume limit in the string-breaking phase, explicit 
centre symmetry breaking can be neglected implying 
$$ 
p_{nm} \; = \; \frac{1}{N_c^2} \hbo \Rightarrow \hbo \tau \; = \; 
1 \, - \, \frac{1}{N_c} \; . 
$$
In the Higgs or the high temperature phase, the whole lattice
belongs to one centre sector, and we find: 
$$ 
p_{nm} \; = \; \frac{1}{N_c} \, \delta _{nm} \hbo \Rightarrow \hbo \tau \; =
\;  0 \; . 
$$ 
We have numerically estimated the transition coefficient $\tau $ 
using a $24^3 \times 6$ lattice, $\kappa = 0.31$ and $\lambda = 0.1$. 
The parameter setting is such that the theory is in the string-breaking phase
for a $24^4$ lattice size. Figure~\ref{fig:7} shows our findings for 
$\tau $ as a function $\beta$. We do find that for 
$\beta_\mathrm{Wil} \stackrel{<}{_\sim} 2.35$ centre-sector transitions do
occur with  
high probability while {\it one } centre-sector is observed throughout 
the lattice universe for high temperatures, i.e., $\beta_\mathrm{Wil} > 2.35$.

\section{Conclusions}

For SU(N) Yang-Mills in 4 dimensions on a torus, we started to investigate 
the ``empty vacuum'', i.e., the set of configurations for which all 
holonomies calculated for any contractible loop yields the unit element of 
the group. All these configurations produce zero field strength everywhere. 
We found a continuous set of {\it gauge in-equivalent} configurations related 
by a symmetry transformation. Including quantum fluctuations, the degeneracy 
of these states is lifted. The symmetry collapses to the well-known 
$Z_N$ centre symmetry which divides the gluonic configurations into 
centre sectors. Transitions between the centre sectors turned out to 
be the key ingredient for {\it confinement}\footnote{although they do not 
explain the confinement energy scale of several hundred MeVs for QCD}.  
Including dynamical matter 
which transforms under the fundamental representation of the group, this 
centre symmetry is {\it explicitly broken}. 
Our central working hypothesis to start with was that transitions between 
centre sectors still take place in the so-called {\it hadronic} phase and 
that these transitions only cease to exist at high temperatures when the 
centre symmetry is also spontaneously broken (on top of the explicit
breaking). 

\vskip 0.3cm 
Before corroborating this picture in the sections~\ref{sec:gi} and
\ref{sec:higgs}, we studied the phenomenological impact in the
Schwinger-model, since it allows for explicit analytical solutions, and 
in an SU(3) quark model for the sake of its relevance for QCD. 
For an even number of colours, it was firstly pointed out 
in ~\cite{Langfeld:2009cd,Langfeldect2010} that quarks acquire {\it 
periodic boundary conditions} in some of the centre sectors. 
For a finite chemical potential, the quarks then might undergo condensation 
due to Bose statistics. In analogy, this has been called {\it 
Fermi Einstein condensation} (FEC). We traced out the roots of FEC 
in the Schwinger model on the torus as finite chemical potential. 
We found that the phenomenological importance of the centre transitions 
is the solution of the {\it silver blaze problem}: centre transitions wipe 
any dependence of the partition function on the chemical potential 
as it must be since the physical states of the model carry no net 
baryon number. 

\vskip 0.3cm 
We then studied an effective SU(3) quark model with constituent quark mass 
$m$ which only interactions are with the gluonic background field 
specifying the centre sector. 
We verified that this very simple model already confines quarks: e.g., 
we considered the thermal energy density as a function of (low) temperatures 
and found that its lowest excitations are mesons of mass $\sim 2m$ 
rather than quarks with mass $\sim m$. By studying the centre sector weights 
(provided by the model), we found that the centre sectors democraticly 
contribute to the partition function in the hadronic phase while under extreme 
conditions basically only the trivial centre sector contributes implying 
that the model merges with the Fermi-gas model in the quark gluon plasma
phase. Using the centre weight of the trivial centre sector as an order 
parameter, we were able to map out the phase diagram of the model as a function 
of the chemical potential and the temperature. We found that FEC is important 
at low temperatures and intermediate values of the chemical potential. 

\vskip 0.3cm 
For FEC to happen in dense QCD, the central question is whether 
centre sector transitions do take place despite of the explicit breaking 
by matter fields. This question can be studied for vanishing chemical 
potential where Monte-Carlo simulations are readily available. 
Using a 3-dimensional $\Z_2$ gauge theory with Ising matter, 
we firstly studied the tension of a centre interface when the model 
is forced into the trivial centre sector (i.e., the sector for which 
the model is identical to the standard ferromagnetic Ising model). 
While the interface tension vanishes in the infinite-volume zero-temperature 
limit, it is finite at high temperatures when centre symmetry is spontaneously 
broken. Giving up the artificial constraint which ties the $\Z_2$ gauge theory 
with matter to the standard Ising model, we find that 
interface tension vanishes for any finite volume. 
Nevertheless, a detailed study of the Polyakov line expectation value 
reveals at rather small value (though non-zero) in the would-be 
confinement phase, and large values in the Higgs phase. 

\vskip 0.3cm 
A theory which is more relevant for QCD, is the SU(2) gauge theory 
with dynamical matter. We have chosen a scalar Higgs field since 
it allows large statistic Monte-Carlo simulations (facilitated by
e.g.~the generalised Luescher-Weisz method) and at the same it explicitly
breaks centre symmetry as quarks do. We firstly identified the region 
of the coupling space where the theory realises string breaking 
well within the size of our lattice. We then studied the extent of 
{\it explicit } centre symmetry breaking by calculating the Polyakov line 
distribution function. At high temperatures, the explicit breaking 
is largely amplified by an additional spontaneous breakdown. 
A study of the temperature dependence of the Polyakov line 
showed quite a similar behaviour as in the Schwinger model. This renders 
cumbersome any conclusion on the spontaneous centre symmetry breakdown 
since it is clearly absent in the Schwinger model. 
This calls for a new type of order parameter which is designed to be 
a Litmus paper for centre transitions. For such an order parameter, we 
here proposed to map the Polyakov line to the centre sector 
for each half of the spatial lattice universe separately, and ask 
for the probability that a particular lattice configuration belongs to {\it 
different} centre sectors in each half. If this is the case, the 
theory certainly undergoes transitions between the centre sectors. 
Using this order parameter, we find clear numerical evidence in the 
SU(2) Higgs theory for a ``hadronic'' phase at small temperatures and 
a de-confined phase at high temperatures. 

\vskip 0.3cm 
In conclusion, we found evidence that dynamical matter does not 
prevent centre sector transitions in QCD-like theories in the low 
temperature ``hadronic'' phase. These transitions only cease to exist 
at high temperatures when centre symmetry is also broken spontaneously. 
We highlighted the phenomenological impact of these centre transitions: 
in the Schwinger model, they solve the Silver-Blaze problem, and 
in an SU(3) quark model they lead to a new phase featuring 
Fermi Einstein condensation (FEC) for cold but dense matter. 
The FEC mechanism only uses fairly robust assumptions on the 
realisation of centre symmetry. Hence, FEC might also be at work 
in QCD  at low temperatures and intermediate values of the chemical potential 
for which quarks are not yet liberated.

\bigskip 
{\bf Acknowledgments: } 
The calculations have been carried out within the DiRAC framework 
at the {\tt High Performance Computing Centre} at University of Plymouth. 
We are indebted to the staff for support. This work is a project of the 
UKQCD collaboration. DiRAC is supported by STFC. 

We would like to thank Philippe de Forcrand, Simon Hands, Emil Mottola, 
Lorenz von Smekal, Bj{\"o}rn Wellegehausen and Axel Maas 
for helpful discussions.

\appendix 
\section{The Schwinger-model calculations \label{app:schwinger} }
In this appendix we extend known results about the Schwinger model
at finite temperature $T=1/\beta$ \cite{Sachs:1992pa} to the case with chemical 
potential $\mu\neq 0$. For related results on the Thirring model see \cite{Sachs:1995dm}.
We enclose the system in an interval with length $L$ and impose periodic
boundary conditions in the spatial direction.

\subsection{Schwinger-proper time regularisation \label{app:swp}}
Any gauge potential with vanishing instanton number on the two-dimensional 
torus can be decomposed as
\be
A_0=-\partial_1\phi+\partial_0\lambda+\frac{2\pi}{\beta}h_0\quad,\quad
A_1=\partial_0\phi+\partial_1\lambda+\frac{2\pi}{L}h_1
\label{eq:k1a}
\en
with constant toron fields $h_0$ and $h_1$. The determinant of 
the Dirac operator for massless fermions does not depend on $\lambda$ 
and its $\phi$-dependence follows from the axial anomaly \cite{Sachs:1992pa}. 
Here we calculate the determinant for arbitrary toron
fields $h_\mu$ and for a chemical potential $\mu\neq 0$.
For constant fields the Dirac operator possesses plane waves as eigenfunctions and
its determinant can be written as an infinite product over all 
admitted momenta,
\be
\det(\I\fdirac_{h,\mu}) = \prod _{(m,n)\in \Z^2} D_{h,\mu}\left(m,E_n\right) \, ,
\label{eq:k3} 
\en
The eigenfunctions are anti-periodic in time and periodic in
space such that
\be
D_{h,\mu}(m,E_n) =
\left(\frac{2\pi }{\beta }\right)^2 \left(m + \ft{1}{2}-\gamma \right)^2 
+E_n^2,\quad \gamma=h_0+\frac{\I\beta}{2\pi}\mu,
\label{eq:k4}
\en
where $LE_n=2\pi\vert n-h_1\vert$.
We introduce the fermionic effective action $\Gamma_{h,\mu}$ by 
\be 
\Gamma_{h,\mu}= \ln \det(\I\fdirac_{h,\mu})
\label{eq:k5}
\en
and obtain in Schwinger proper time regularisation 
\be
\Gamma_\Lambda(\mu, h)=\int _{1/\Lambda ^2} ^\infty
\frac{ds}{s} \; \sum _{m,n} \; 
\E^{- s D_{h,\mu}(m,E_n)}\equiv
\sum_{n} \Gamma_\mathrm{fer}(E_n,h,\mu)\;.
\label{eq:mu1}
\en
The terms in the last sum are given by the integral
\be
\Gamma_\mathrm{fer}(E,h,\mu) =  2\int _E ^\infty de \, e\,  W(e) \;,
\label{eq:mu4}
\en
where the integrand contains the function
\be
W(E) = \int_{1/\Lambda ^2} ^\infty
ds \; \sum _{m}  \; \exp \left\{ - s D_{h,\mu}(m,E)\right\}\;. . 
\label{eq:mu3}
\en
Now we can do the proper time integration. Since the resulting sum over $m$ is
convergent we may set $T/\Lambda=0$ such that
\be
W(E)  = \E^{-E^2/\Lambda^2} \sum_{m} f(m,E)\quad\hbox{with}\quad
f(m,E)=\frac{1}{D_{h,\mu}(m,E) } \;  .
\label{eq:mu5}
\en
The sum over $m$ can be rewritten
with the help of the Poisson resummation formula
$$
\sum _n f(m,E)  =  \sum _k \tilde f(k,E) \quad ,\quad
\tilde f(k,E)  =  \int dx \; \mathrm{e}^{ 2 \pi\I k  x } \; f(x,E) \; .
$$
After a shift of the integration variable the last integral 
takes the form
\be
\tilde f(k,E) =\frac{(-1)^k}{2\pi}\beta^2\mathrm{e}^{2\pi\I k h_0} \int dx \;
\frac{ \mathrm{e}^{i k  x } }
{(x-\I\beta\mu)^2+(\beta E)^2} \;  .
\label{eq:mu6}
\en
The integral over $x$ is known in closed form and the resulting sum over $k$ 
can easily be computed and yields 
\be
\sum_{k} \tilde f(k, E) =\frac{1}{2E}\frac{d}{dE}
\left\{\beta E
-\log\left(1+\E^{ 2\pi\I\gamma-\beta E}\right)
-\log\left(1+\E^{-2\pi\I\gamma-\beta E}\right)\right\}\;.
\label{eq:mu7}
\en
Multiplied with $\exp(-E^2/\Lambda^2)$ this becomes the function
$W(E)$ in (\ref{eq:mu4}). The $\mu$-dependent part is UV-finite and 
hence we can safely remove the  cutoff there leaving us with 
\be 
\Gamma_{\Lambda}(\beta,\mu, h) =  \beta E_{\Lambda}(L,h_1)
+\Gamma_1(\beta,L,\mu,h)\;,\label{eq:k7}
\en
with a divergent zero-point energy $E_\Lambda$
and a finite temperature correction
\be
\Gamma_1(\beta,L,\mu,h)= \sum _{n} 
\ln \Big\{\left( 1 \;+\;  \E^{2\pi\I h_0} \, 
\E^{- \beta (E_n + \mu)} \right)
+ 
\ln \left( 1  +  \E^{- 2\pi\I h_0} 
\,\E^{ - \beta (E_n - \mu) } \right)\Big\} \; . 
\label{eq:kb7b} 
\en
In the zero-temperature limit we must recover the well-known
Casimir energy of fermions on a circle with circumference $L$
\cite{Bordag:2001qi,Kiefer:1993fw},
\be
\Gamma_\Lambda(\beta,L,\mu,h)\stackrel{\beta\to\infty}{\longrightarrow}
-\beta E_{\rm Cas},\quad 
E_{\rm  Cas}=-\frac{\pi}{6L}+\frac{2\pi}{L}\left(\ha-h_1\right)^2.
\label{casimir} 
\en
The Casimir energy is periodic in $h_1$ with period $1$ and
in the last formula we must assume $h_1\in [0,1]$.
We conclude that the renormalised effective action is
\be
\Gamma(\beta,L,\mu,h)=-\beta E_{\rm Cas}(L,h_1)+\Gamma_1(\beta,L,\mu,h)\,.\label{eq:k8}
\en
Note that the energies $E_n$ are proportional to $1/L$ such that
$\Gamma_1$ depends only via the dimensionless parameter $\tau=\beta/L$ 
on the size $L$ of the system. 

\subsection{$\theta$-function representation and integration
over toron fields\label{app:theta} }
The effective action $\Gamma=\log\det(\I\fdirac)$
can be expressed in terms of the $\theta$- and $\eta$-function \cite{Sachs:1995dm}
\be
\det(\I\fdirac_{h,\mu})=\frac{1}{ \eta^2(\I\tau)}
\Theta\genfrac{[}{]}{0pt}{}{h_1-\ha}{\gamma}(0,\I\tau)\;
\Theta\genfrac{[}{]}{0pt}{}{h_1-\ha}{-\gamma}(0,\I\tau)\;,
\label{app1}
\en
where $\gamma$ is defined in (\ref{eq:k4}) and we used the Dedekind
eta-function 
\be
\eta(\I\tau)= \E^{-\pi\tau/12}\prod_{n>0} \left(1-\E^{-2\pi \tau n}\right),\quad
\tau=\frac{\beta}{L}\;,
\label{app7}
\en
and the theta function \cite{mumford}
\be
\Theta\genfrac{[}{]}{0pt}{}{\alpha}{\gamma}(0,\I\tau)=
\sum _{n\in\Z} \E^{-\pi\tau (n+\alpha)^2+2\pi \I(n+\alpha)\gamma}\label{app2}
\en
which has the product expansion
\begin{eqnarray}
\frac{1}{\eta(\I\tau)}\Theta\genfrac{[}{]}{0pt}{}{\alpha}{\gamma}(0,\I\tau)&=&
\E^{2\pi\I\alpha\gamma} \E^{-\pi \tau\alpha^2+\pi \tau/12}\nonumber\\
&&\hskip-10mm\cdot\prod_{n=1}^\infty \left(1+\E^{-2\pi \tau(n+\alpha-1/2)}\E^{2\pi\I\gamma}\right)
 \left(1+\E^{-2\pi \tau(n-\alpha-1/2)}\E^{-2\pi\I\gamma}\right)\;.\label{app5}
\end{eqnarray} 
Note that the determinant is invariant under large gauge 
transformations $h_\mu\to h_\mu+1$.

\vskip 0.3cm 
Using the product expansion for the $\theta$-function it follows at once 
that the effective action 
$\Gamma(\beta,L,\mu,h)=\log \det(\I\fdirac_{h,\mu})$ has the series
expansion (\ref{eq:k8}).
In the derivation we assumed that $h_1$ takes its values in the unit interval
and this is why the $h_1$-periodicity of the determinant is not
manifest. In the thermodynamic limit we find the following expression for
the free energy density $f$ in $Z=\exp(-\beta L f)$:
\begin{eqnarray}
f\stackrel{L\to\infty}{\longrightarrow}-\frac{1}{\pi\beta}\int_{0}^\infty  
dp \; \log \left[ \left(1+\E^{2\pi \I h_0}\E^{-\beta (p+\mu)}\right) 
\left(1+\E^{-2\pi \I h_0}\E^{-\beta (p-\mu)}\right) \right] 
\nonumber 
\\
=\frac{1}{\pi\beta^2}\left[
\hbox{dilog}\left(1+\E^{2\pi \I
  h_0-\beta\mu}\right)+\hbox{dilog}\left(1+\E^{-2\pi \I
  h_0+\beta\mu}\right)\right]. 
\label{freeenergy1} 
\end{eqnarray}
For $L/\beta\to\infty$ the free energy density does not depend on the
constant gauge field $h_1$, as expected. 
When one varies the constant gauge field $h_0$ (or equivalently the boundary
condition in the temporal direction) then one smoothly interpolates between
free fermions and free bosons or between a Fermi-Dirac and a Bose-Einstein
distribution.

In a gauge theory we must average over all gauge fields
and in particular we must integrate the fermionic determinant
over the constant gauge fields as well. The integral of (\ref{app1}) over $h_1$
can be done explicitly. If $m,n$ are the summation indices in the
double sum (\ref{app1}) we change summation indices according to
$m=p+q$ and $n=q$. The sum over $q$ together with the $h_1$-integral 
over $[0,1]$ turns into an integral over $[-\infty,\infty]$ and 
yields the simple result (\ref{eq:kk10}).
The final integration over $h_0$ is easily performed and yields
the $\mu$-independent result (\ref{eq:kk11}) for the averaged
determinant $\int dh_0 dh_1\det(\I\fdirac_{h,\mu})$.

\subsection{Polyakov loops \label{app:pol}}
Let us finally calculate the expectation values of product of Polyakov loops
\be
P_q(u)=\E^{iq\int dx^0 eA_0(x^0,u)}=\E^{2\pi \I qh_0}\E^{-\I eq\int \partial_u\phi(x^0,u)dx^0}
\en
corresponding to static charges $q\in\Z$. In particular $P_{-q}=\bar P_q$.
First we do the integration over the harmonics. With the help of (\ref{eq:kk10}) 
we obtain
\begin{eqnarray*}
\int dh_0dh_1 \det\left(\I\slashed{\partial}_{h,\mu}\right) \prod P_{q_i}(u_i)=
\frac{1}{\sqrt{2\tau}}\frac{1}{\eta^2(\I\tau)}\E^{-\pi\tau\,Q^2/2
+\beta\mu\, Q}\E^{-\I \int d^2x\, j(x) \phi(x)}\label{polw1}
\end{eqnarray*}
with total charge $Q=\sum q_i$ and source
\be
j(x)=e\sum_i q_i \delta'(x^1-u_i).\label{polw3}
\en
The functional integral over $\phi$ has been done previously
in \cite{Sachs:1992pa} and yields
\be
\left\langle\prod P_{q_i}(u_i)\right\rangle=
\prod_i \left\langle P_{q_i}\right\rangle
\E^{-\beta\sum_{i\neq j} q_i q_j V(\vert u_i-u_j\vert)}\label{polw9}
\en
with expectation values (\ref{eq:kk20}) for the individual Polyakov loops
and periodic potential
\be
V(\vert u\vert)=\frac{\pi m_\gamma}{4}
\frac{\cosh\left(\frac{m_\gamma}{2}(L-2\vert u\vert)\right)}
{\sinh\left(\frac{m_\gamma L}{2}\right)},\quad \vert u\vert\leq L\;.\label{polw13}
\en
The  potential energy decreases exponentially fast for large separations of
the charges
\be
V(\vert u\vert)\sim \frac{\pi m_\gamma}{4} \,\E^{-m_\gamma L/2}\;.
\label{polw15}
\en

\section{Local-hybrid Monte Carlo for the Higgs sector} 

For the update the scalar field in accordance to the functional integral 
(\ref{eq:qv10}), we here discuss the Local Hybrid-MC (LHMC) scheme. 
Assume that $\phi (x)$ is chosen for the update. 
The LHMC Hamiltonian is given by 
\be 
H \; = \; \frac{1}{2} \, 
\pi ^\dagger \pi \; + \; S_L(\phi, \phi ^\dagger) . 
\label{eq:a1}
\en
To set up the LHMC scheme, it is convenient to work with 
real variables, i.e., $\pi = R + i S$, $\phi = r + i s$, $s,S,r,R \in \R$.
The Hamilton-Jacobi equations of motion are given by 
\bea 
\dot{r} &=& \frac{\partial H}{\partial R} \; = \; R \, , \hbo 
\dot{s} \; = \; \frac{\partial H}{\partial S} \; = \; S \, , 
\label{eq:a2} \\ 
\dot{R}  &=& - \, \frac{\partial S_L}{\partial r} \; = \; - 
\frac{\partial S_L}{\partial \phi} - \frac{\partial S_L}{\partial
  \phi^\dagger} ,  \hbo 
\dot{S}  \; = \;  - \, \frac{\partial S_L}{\partial s} \; = \; 
- \I \Bigl[ \frac{\partial S_L}{\partial \phi} -   \frac{\partial S_L}{\partial
   \phi^\dagger} \Bigr] \; . 
\label{eq:a3} 
\ena 
It is easy to check that the above equations imply $\dot{H}=0$. 
Combining both equations in (\ref{eq:a3}) yields: 
\be 
\dot{\pi} \; = \; \dot{R} + \I \dot{S} \; = \; -\, 2 \, \frac{\partial
  S_L}{\partial \phi^\dagger}  , \hbo 
\dot{\phi} \; = \; \dot{r} + \I\, \dot{s} \; = \; \pi \; . 
\label{eq:a4} 
\en
For the action in (\ref{eq:qv10}), the terms of $S_\mathrm{Higgs}$ which 
depend on $\phi (x)$ or $\phi ^\dagger (x)$ give rise to 
\bea 
S_L &=& - \frac{\kappa }{2} \, \sum _{\mu} \Bigl[ 
\phi ^\dagger (x) \, U_\mu (x) \, \phi (x+\mu) \, + \, 
\phi ^\dagger (x) \, U^\dagger_\mu (x-\mu) \, \phi (x-\mu) \, \Bigr] 
\nonumber \\ 
&-& \frac{\kappa }{2} \, \sum _{\mu} \Bigl[ 
\phi ^\dagger (x-\mu) \, U_\mu (x-\mu) \, \phi (x) \, + \, 
\phi ^\dagger (x+\mu) \, U^\dagger_\mu (x) \, \phi (x) \, \Bigr] 
\nonumber \\ 
&+& \frac{1}{2} \, \phi^\dagger (x) \, \phi(x) \; + \; 
\lambda \; [\phi^\dagger (x) \, \phi(x) ]^2 \; . 
\nonumber
\ena
Hence, defining 
\be 
B(x) \; = \; \sum _\mu \Bigl[ \, U_\mu (x) \, \phi (x+\mu) \, + \, 
 U^\dagger_\mu (x-\mu) \, \phi (x-\mu) \Bigr] \; , 
\label{eq:a5} 
\en
we finally find: 
\be 
\dot{\phi } \; = \; \pi \; , \hbo 
\dot{\pi} \; = \; \kappa \, B(x) \; - \; \phi(x) \, \Bigl[ 1 + 4 \,  \lambda 
\, \phi^\dagger (x) \phi (x) \, \Bigr] \; . 
\label{eq:a6} 
\en  

\end{document}